\PassOptionsToPackage{fleqn}{amsmath} %
\documentclass[9pt,sigconf,review=false, anonymous=false, 
    timestamp=false, balance=false, nonacm=true, screen=true, authordraft=false]{acmart}		%

\input{datalabmacros}

\setcopyright{cc} %

\makeatletter
\renewcommand\@copyrightpermission{%
  \begingroup\footnotesize\noindent
  This paper is published under the Creative Commons Attribution 4.0 International
  (CC-BY 4.0) license. Authors reserve their rights to disseminate the work on their
  personal and corporate Web sites with the appropriate attribution, provided that you
  attribute the original work to the authors and CIDR 2026. 16th Annual Conference on
  Innovative Data Systems Research (CIDR '26). January 18-21, 2026, Chaminade, USA.
  \par\endgroup
}
\makeatother

\begin{document}

\newcommand{\ourTitle}{Database Research needs an Abstract Relational Query Language}
\title[\ourTitle]{\ourTitle}

\author{Wolfgang Gatterbauer}
\orcid{0000-0002-9614-0504}
\affiliation{%
    \orcidicon{0000-0002-9614-0504}
	Northeastern University\country{USA}}

\author{Diandre Miguel B. Sabale}
\orcid{0009-0005-7689-1756}%
\affiliation{%
    \orcidicon{0009-0005-7689-1756}
	Northeastern University\country{USA}}


\begin{abstract}
For decades, \SQL has been the default language for composing queries, 
but it is increasingly used as an artifact to be read and verified rather than authored.
With Large Language Models (LLMs), queries are increasingly machine-generated, while humans read, validate, and debug them.
This shift turns relational query languages into interfaces for back-and-forth \emph{communication about intent},
which will lead to a rethinking of relational language design,
and more broadly, relational interface design.

We argue that this rethinking needs support from an
\emph{Abstract Relational Query Language} (ARQL):
a semantics-first reference metalanguage
that separates query intent from user-facing syntax
and makes underlying relational patterns explicit and comparable across user-facing languages. 
An ARQL separates a query into 
($i$) a \emph{relational core} (the compositional structure that determines intent), 
($ii$) \emph{modalities} (alternative representations of that core tailored to different audiences), 
and ($iii$) conventions (orthogonal environment-level semantic parameters under which the core is interpreted,
e.g., set vs.\ bag semantics, or treatment of null values). 
Usability for humans or machines then depends less on choosing a particular language and more on choosing an appropriate modality.
Comparing languages becomes a question of which relational patterns they support and what conventions they choose.

We introduce Abstract Relational Calculus (\ARC), 
a strict generalization of Tuple Relational Calculus (\TRC), 
as a concrete instance of ARQL. 
\ARC\ comes in three \HL{modalities}[modality]: 
($i$) a comprehension-style textual notation, 
($ii$) an Abstract Language Tree (ALT) for machine reasoning about meaning, 
and ($iii$) a diagrammatic hierarchical-graph (higraph) representation for humans. 
\ARC\ provides the missing vocabulary and acts as a Rosetta Stone for relational querying.
Because its ALT makes binding, scoping, and grouping structure explicit in a small, reusable operator vocabulary, it is a natural intermediate target for automated query generation and validation, including NL2SQL systems that translate natural language into a semantic query representation and then render it to \SQL.

\end{abstract}

\maketitle

\noindent
\raisebox{-.4ex}{\HandRight}\hspace{.2cm}This PDF contains internal hyperlinks for easier reading: click any 
{\HL{linked term}[notion-ex]}
to jump to 
\HA{the section where it is defined}[notion-ex].

\section{Introduction}
\label{sec:introduction}

\introparagraph{New interfaces for humans and machines}
Several recent efforts question \SQL as the 
default
relational query language (QL) and propose to either extend or completely replace it
\cite{DBLP:conf/cidr/0001L24,
DBLP:journals/pvldb/ShuteBBBDKLMMSWWY24,
DBLP:conf/sigmod/ArefGKLMMMMNPRS25}.
Examples in these debates include whether
nested correlated queries are inherently hard for users to follow and should be replaced with more dataflow (algebraic) abstractions, 
and whether set or bag semantics are the right choice 
(e.g., debated at the DBPL workshop at SIGMOD'25~\cite{DBPL:2025}).
Less explored, however, is a more basic representational question: 
\circled{1}
\emph{How should we represent the intent of a query so that its relational structure is explicit and comparable across different surface syntaxes? How can we describe how a query composes its inputs (i.e., the base relations) to define query intent, independent of the idiosyncrasies of any particular query language?}

\begin{figure}[t]
    \centering        
    \includegraphics[scale=0.36]{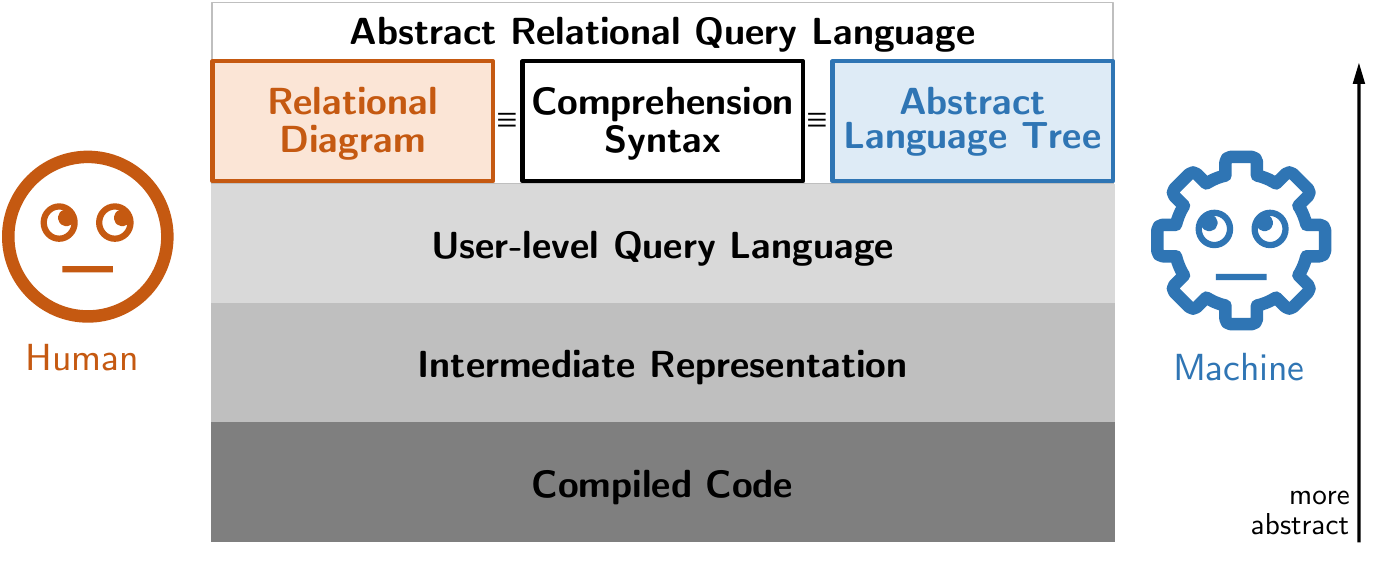}
    \caption{An \emph{Abstract Relational QL} (ARQL) abstracts away from syntactic details of a query 
    to a higher-level representation.
    Just as Intermediate Representations (IRs) enable query optimization, 
    a more abstract representation can support semantic understanding of a query's intent.
    Both humans and machines can benefit from \emph{\HL{modalities}[modality]} tailored to their needs.
    \emph{\HL{Conventions}[convention]} (not shown) factor out orthogonal design choices that don't affect the relational pattern.}
    \label{Fig_Layers_of_Abstraction}
\end{figure}

At the same time, relational queries are increasingly produced by machines and validated by humans. 
In this setting, the user interaction with relational databases changes:
now \SQL is not just a language for users composing queries, but it also increasingly serves as a message format between machine generation and human validation.
As noted in the Cambridge report~\cite{DBLP:journals/corr/abs-2504-11259}, these developments ``\emph{potentially change how we interface with relational databases},'' 
shifting emphasis from query composition to query interpretation. 
``\emph{The essential skill is no longer simply writing programs but learning to read, understand, critique and improve them instead}''~\cite{NYTimes:CS:dont:panic}.
Since LLMs can hallucinate or introduce errors, 
``\emph{effective explanation mechanisms ... become increasingly important}''~\cite{DBLP:journals/corr/abs-2504-11259}. 
This raises a second question: 
\circled{2}
\emph{How should machine-generated queries be presented to users so they can validate them and provide  feedback?}

The challenge is not limited to human-facing interfaces. 
Machine-facing tasks such as semantic similarity search and retrieval also require representations aligned with meaning rather than syntax. 
\SQL's surface syntax is a poor proxy for intent: 
semantically equivalent queries can differ substantially in syntactic structure, while syntactically similar queries may encode different semantics. 
In the NL2SQL domain, current benchmarks often rely on surface-level criteria such as exact string match
or execution match.
Since those fail to capture deeper semantic relationships,
Floratou et al.~\cite{DBLP:conf/cidr/FloratouPZDHTCA24} argue for ``\emph{a shift towards intent-based benchmarking frameworks}.''
This raises a third question: 
\circled{3} 
\emph{What language abstraction should an LLM (or future machine-tool) 
use to internally reason about query intent and semantic similarity in a way that is faithful to relational meaning?}

\introparagraph{Our Suggestion}
We believe that database research needs \emph{new vocabulary} to analyze relational intent across languages
and a \emph{semantics-focused representation} of relational languages 
that decouples intent from user-facing syntax while supporting multiple modalities
(i.e., mechanically inter-translatable 
representations of the same language tailored to different audiences, without treating each modality as a separate language).
We call such a representation an \HA{\emph{Abstract Relational Query Language} (ARQL)}[ARQL].

It is \emph{abstract} because it factors out differences in concrete syntax and design choices 
and instead focuses on a small set of 
compositional relational primitives shared across relational query languages.
In this sense, a representation is \emph{more abstract} 
when it makes the relational intent of a query~\cite{gatterbauer2011databases,10.1145/1951365.1951440}
explicit without relying on syntactic shortcuts
or being forced to expose ``conventions'' in the language. 
In general, we refer to a language-agnostic description of how data is transformed from input to output as the \emph{relational pattern} of a query~\cite{DBLP:journals/sigmod/GatterbauerD25}.
Our goal is a clean \emph{separation of concerns}.
Just as Intermediate Representations (IRs),
such as SDQL~\cite{DBLP:journals/pacmpl/ShaikhhaHSO22} and Substrait~\cite{substrait},
decouple front-end parsing from optimization and code generation,
we want to enable a \emph{syntax-agnostic discussion of language features
at a more conceptual level}.
This lets us treat relational patterns as modules (\cref{sec:abstract relations})
and
several issues that are not necessarily part of the relational pattern of a query 
as orthogonal choices (``conventions'', see \cref{sec:conventions}),
such as the convention of using set or bag semantics, 
the blurry distinction between declarative and procedural languages, 
the treatment of null values,
typing and casting conventions,
as well as different initializations of aggregate functions (e.g., 0 or null for \texttt{sum}).
It also allows us to discuss 
the many syntactic variants that SQL permits for expressing basically the same intent,
as well as when rewrites are not equivalent (e.g., the COUNT bug, see~\cref{sec:count bug}).

Another concrete use case is NL2SQL. Rather than generating SQL text directly, an NL2SQL system can generate an ARQL representation of intent and then render it into SQL. 
In this paper, our concrete ARQL instance is Abstract Relational Calculus (\ARC) (introduced in \cref{sec: ARC}) with a machine-facing Abstract Language Tree (ALT) modality, which provides a natural intermediate target for NL2SQL systems.

\introparagraph{Modalities instead of languages}
An ARQL does not have just one representation.
Instead, we propose developing alternative \emph{modalities} of the same language, each tailored to different purposes.
These modalities offer alternative views of a query, 
targeted for either human interpretation or machine reasoning.

Concretely, what are usually called \emph{Abstract Syntax Trees} (ASTs) 
tend to remain too close to the concrete syntax of a language. 
For example, SQLGlot's AST~\cite{sqlglot:view,tobiko}
places JOINs as children of SELECT nodes, 
which reflects 
surface-level parsing (concrete syntax) rather
than abstract semantic relationships. 
We argue that such representations fall short as truly abstract representations of a query.

We use the term \HA{\emph{Abstract Language Tree} (ALT)}[ALT]
for a universal, hierarchically structured representation of the \emph{semantics of a query} rather than its 
syntax.
We originally used the term \emph{Abstract Language Higraph} (ALH) instead of ALT. 
The motivation is that ASTs and ALTs
are ``trees'' only with respect to the containment (nesting) structure
(and even that can be blurred by correlated constructs such as \texttt{LATERAL} joins).
Once name resolution is performed and bindings are established
(i.e., identifier occurrences are connected to their declarations via cross-references, 
and the resulting structure is often called an \emph{annotated/decorated AST}~\cite{Cooper:Torczon:Engineering:a:compiler:3}),
the overall structure is better viewed as a hierarchical graph (a tree of containment with additional edges). 
This also matches the intuition that lexical scopes correspond to nested regions (\cref{Fig_simple_TRC_RD}). 
\emph{Higraphs}~\cite{DBLP:journals/cacm/Harel88}  
formalize exactly this combination of nesting and linking:
nodes may be nested within nodes (capturing containment/scopes) while edges capture references. 
However, they are not widely known, 
and the term is unfamiliar to many readers
(for a simplified and accessible higraph formalism, see the online appendix of \cite{disjunctions}).
Thus, we ultimately kept the simpler term Abstract Language Tree (ALT): 
the conceptual shift from surface syntax to the underlying semantic operations is already substantial, 
while the remaining intuitions from ASTs carry over,
and our higraph modality continues to make the hierarchical graph (higraph) data structure explicit.

Similar to how query graphs support the optimization of conjunctive queries~\cite{moerkotte:compilers}, 
we believe ALTs provide a better data structure for semantic analysis of relational queries.
For an ARQL, the language-independent ALT is ideally identical to its AST, because the syntax reflects its semantics.

Importantly, ALTs can also be rendered diagrammatically for human users
as hierarchical graphs (higraphs). In that form, the nested scopes of nodes in the ALT
are replaced by a nesting of nodes.
Prior user studies
have shown that \diagrams{} can help users understand relational structures 
faster and more reliably~\cite{Leventidis2020QueryVis, 10.1145/3639316}.
This emphasizes a key distinction: \emph{language design} should not be conflated with \emph{interface usability}. 
Whether users ``like'' a language is a question of modality, not of the language core itself. 
Instead, modalities should be designed with target consumers in mind, 
i.e.\ human-facing modalities for accurate semantic understanding and debugging, and machine-facing modalities for tasks such as semantic similarity assessment or query transformation.
Thus, an ARQL provides the relational structure, while modalities are lossless presentations of that structure for different consumers.

While we agree with the observation that ``\emph{the idea of a single, universal language or paradigm ... covering all data programming needs is unlikely}''~\cite{DBLP:journals/corr/abs-2504-11259}, 
we argue that many of these needs can be addressed at the level of modalities instead of languages.
The goal of an ARQL is not to unify all QLs under a single syntax, 
but to enable meaningful comparisons across languages in terms of their underlying relational patterns,
and different modalities can serve the respective needs of humans and machines.
The translation between modalities can also be automated.

\introparagraph{Conventions instead of languages}
In Green's cognitive dimensions of notations~\cite{10.5555/92968.93015}
a ``system'' consists of notation (the representational form)
and an environment (the surrounding tooling).
Similarly, we suggest distinguishing between \emph{a language}
(a representation that encodes the relational composition of a query in a particular surface syntax)
and a \emph{convention} (an orthogonal design decision that can be switched 
and will affect the behavior
but not the relational core).
For example, the aggregate \texttt{sum(R.A)} initializes with null in \SQL, but with 0 in Souffl\'e (\cref{sec:conventions}).
This difference is a design decision.
It is a convention that does not affect the way a relational query composes its various components to encode a meaning.
An ARQL focuses only on the relational patterns of a query and does not expose conventions which are specified separately in the surrounding environment.
With this change, a sufficiently generic language design could be interpreted under either set or bag semantics.
It is just a switch that we flip on or off.
While discussion of set vs.\ bag semantics 
is still important for query optimization, it becomes orthogonal to ``\emph{language design}.''

\introparagraph{Contributions}
\circled{1}~We suggest that the database community develops abstract representations of relational queries
that can embed relational query patterns across user-facing relational query languages. 
This effort can support, but is orthogonal to, the development of concrete user-level Query Languages (QLs) 
and efforts on Intermediate Representations (IRs) 
(\cref{sec:introduction}).
\circled{2}~We propose a concrete formulation of such a language 
called Abstract Relational Calculus (\ARC),
which is a strict generalization of Tuple Relational Calculus (\TRC) that assumes flat relational inputs and outputs and so far has 3 modalities.
By making implicit relational constructs and dependencies explicit, 
\ARC\ abstracts and surfaces common query patterns found across different relational languages in a more explicit representation.
By treating human- and machine-facing representations as modalities of the same underlying calculus, 
it supports a more principled discussion of relational language design, 
for both the future human and machine audiences (\cref{sec: ARC}).
\circled{3}~We show \ARC\ representations of running examples from recent and older papers, 
which we believe support an ongoing discussion
(\cref{sec:use cases} and examples interspersed throughout \cref{sec: ARC}).

\section{Abstract Relational Calculus (\ARC)}
\label{sec: ARC}

We formalize Abstract Relational Calculus (\ARC), a strict generalization of Tuple Relational Calculus (\TRC)
that models relational query languages in a collection framework.\footnote{We were considering the more explicit name \emph{Abstract Tuple Relational Calculus} to emphasize the lineage and leave space for a possible future Abstract Domain Relational Calculus. 
But the more we thought about it, the more we came to believe that the domain perspective 
is not well-suited for an ARQL
(though it may well be a suitable choice for a user-facing syntax
as in Rel~\cite{DBLP:conf/sigmod/ArefGKLMMMMNPRS25}).
Alternative names considered were Generalized \TRC\ and Extended \TRC\ (as in extended relational algebra).}  

\ARC\ is an Abstract Relational Query Language (ARQL):
a semantics-first reference metalanguage that can encode the core relational query patterns of \SQL\ and various proposed alternatives.
We present \ARC\ in 3 modalities:
($i$) a comprehension-based syntax that generalizes \TRC, 
($ii$) an Abstract Language Tree (ALT) suited for machine reasoning, and 
($iii$) a diagrammatic higraph modality suited for human inspection.
Although equivalent, each modality is tailored to a different audience.

\subsection{Starting with \TRC}

We start with \TRC\ because we have a strong conviction that the named calculus perspective is a more suitable abstraction for an ARQL than positional addressing.
Codd~\cite{DBLP:journals/cacm/Codd82} proposed to 
``\emph{replace positional addressing by totally associative addressing}'',
i.e.\ accessing values by named attributes
rather than by argument position. 
This gives us logical independence not only from tuple order (row position), but also from attribute order (column positions).
Moreover, Boolean statements in \TRC\ are always domain independent~\cite{disjunctions} as long as all range variables are bound to relations,
a property that is not widely known and does not hold for \DRC.

\begin{figure}[t]
  \centering
  \begin{subfigure}[b]{.43\linewidth}
  \centering
      \includegraphics[scale=0.45]{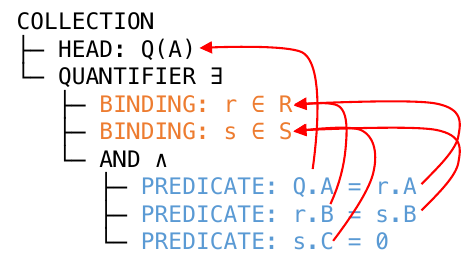}
      \vspace{-3mm}
      \caption{}
      \label{Fig_simple_TRC2}
  \end{subfigure}
	\hspace{1mm}
  \begin{subfigure}[b]{.52\linewidth}
  \centering		
      \vspace{1mm}		
      \includegraphics[scale=0.5]{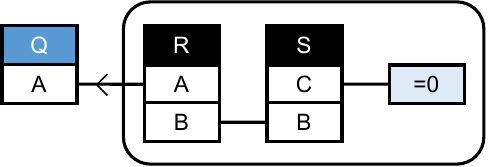}
      \vspace{1mm}
      \caption{}
      \label{Fig_simple_TRC_RD}
  \end{subfigure}
\caption{(a): Linked \HL{Abstract Language Tree (ALT)}[ALT] for \TRC\ \cref{eq:simpleTRC}. The overlaid arrows show the result of the linking step and are conceptual only.
(b): Diagrammatic higraph representation of the linked ALT as a variant of \diagrams.}
\end{figure}

Several recent works in our community are inspired by Datalog, due to its handling of recursion. 
However, nothing prevents us from adding recursion in the named attribute perspective
(\cref{sec:recursion}).
The positional (domain) perspective is also favored for conciseness: 
One can simply write $R(x,y)$ 
for predicate or function application
instead of
$\exists r \in R \,[r.A=x \wedge r.B=y]$.
For an ARQL, however, brevity is diametrically opposed to its goal of surfacing patterns across languages, and making otherwise implicit constructs explicit.
As one example, an \HL{assignment predicate} $Q.A=r.A$ in \TRC,
$\{Q(A) \mid \exists r\in R\,[Q.A=r.A]\}$,
has no explicit counterpart in \DRC,
where the same binding is implicit in the output tuple $\{(x) \mid R(x)\}$.
Conciseness is often associated with usability (fewer letters to type).
But we associate usability primarily with the chosen modality rather than the relational core
(see \cref{Fig_simple_TRC2} vs.\ \cref{Fig_simple_TRC_RD}, which render the same language in two different modalities; the diagrammatic modality supports faster human inspection).

Ignoring notational conventions, 
the following is a valid \TRC\ query according to a widely used textbook~\cite{Elmasri:dq}:
\begin{align*}
  \{ r.A \mid r \in R \wedge \exists s [r.B=s.B \wedge s.C=0 \wedge s \in S] \}
\end{align*}
We make two changes. 
First, we clarify the scopes. Whenever a relation variable is quantified, then it is also bound to a relation:
\begin{align*}
  \{ r.A \mid r \in R, \exists s \in S [r.B=s.B \wedge s.C=0]  \}
\end{align*}
Second, we have stricter scoping rules. \emph{We do not allow variables bound in the body to appear in the head}. 
Instead, we assign values to the head variables explicitly via an \HL{assignment predicate}:
\begin{align}
  \{ \h{Q}(A) \mid \exists r \in R, s\in S[\h{Q.A = r.A} \wedge  r.B=s.B \wedge s.C=0 ] \}
  \label{eq:simpleTRC}
\end{align}
This means that all bindings (e.g., $s\in S$) are now introduced by an explicit quantifier. 
Notice that two bindings can share the same quantifier ($\exists r \in R, s \in S$).
We call the extra predicate $Q.A=r.A$ an \emph{\HA{assignment predicate}} to distinguish it from the other \emph{\HA{comparison predicates}[comparison predicate]}.
This membership-style formalization of \TRC\ is developed in great detail in~\cite{disjunctions}.

\subsection{Language \HA{Modalities}[modality]}

\introparagraph{\HL{Abstract Language Tree (ALT)}[ALT]}
\Cref{Fig_simple_TRC2}
shows our formalism of an Abstract Language Tree (ALT) representation of 
\cref{eq:simpleTRC}
which makes this nesting of one or more bindings under a quantifier explicit~\cite{disjunctions}.
Notice that a query (or a collection) consists of a head and a formula as body,
and a quantification starts the body.

We also show conceptual links from predicates to the bindings of their range variables. 
These are not typically shown in ASTs,
but they reflect the data structures created after the linking step and symbol tables are created.
Given what we perceive as a confusion about 
what an \emph{Abstract} Syntax Tree (AST) is supposed to represent (recall \cref{sec:introduction}), 
we say ALT but actually think about this linked and hierarchical data structure as an 
Abstract Language Higraph (ALH).
Developing a good set of data abstractions 
is essential for solving problems.
We believe a hierarchical graph is a good abstraction for relational structures.

\introparagraph{\diagrams{} (Higraph diagrams)}
For computational analysis, this pointer-based hierarchical graph structure is appropriate. 
For human consumption, we use a diagrammatic representation of the ALT
where scopes represented as nodes in the ALT become regions, and
where the attributes of a table are represented adjacent to the table name instead of using additional edges. 
For the relationally complete fragment, these concepts were already formalized as \diagrams\ in more detail in \cite{10.1145/3639316, disjunctions, DBLP:journals/sigmod/GatterbauerD25}. 
A user study has shown that these diagrams allow humans to recognize and reason about patterns faster than \SQL. 
The user study 
was reproduced~\cite{10.1145/3687998.3717044}.
Two recent tutorials~\cite{ICDE:2024:diagrammatic:tutorial,DBLP:journals/pvldb/Gatterbauer23} give a detailed comparison of this visual formalism against prior work.

Two minor differences from that prior work are: 
($i$) we now explicitly represent existential scopes (previously omitted because, under set semantics, only negation requires an unambiguous scope interpretation; this changes under bag semantics and aggregation), 
and ($ii$) we visually decorate \HL{assignment predicates}[assignment predicate] (crucial for nested comprehensions).

\subsection{Interpreting \TRC\ as set comprehension}
\label{sec:set comprehension}

Everything so far was grounded in first-order logic.
We next interpret relational query languages in a collection framework, 
viewing a query as an expression in a comprehension calculus with tuple variables, quantifiers, and scoping.
This interpretation will allow us later to go beyond first-order logic, yet remain declarative.

A declarative specification of a set can be given by specifying 
the elements that satisfy the properties of the set.
For example, for sets $X$ and $Y$, the subset of their product where the former is smaller than the latter is the set
$\{ (x, y) \mid x < y \wedge x \in X \wedge y \in Y \}$.
With our stricter scoping rules, we would write it instead as
\begin{align*}
\{ (a,b) \mid  \exists x \in X, y \in Y [x<y \wedge a = x \wedge b = y]\}
\end{align*}
In a Haskell list comprehension syntax, this would be written as\footnote{Since value variables must start lowercase in Haskell, we use $xs$ and $ys$ instead of our otherwise preferred notation $X$ and $Y$.}
\begin{align*}
&\texttt{[(x,y) | x <- xs, y <- ys, x < y]}
\end{align*}
with its semantics given by the conceptual evaluation strategy:
\begin{align*}
&\texttt{for x in X:}\\[-1mm]
&\hspace{3mm} \texttt{for y in Y:}\\[-1mm]
&\hspace{6mm} \texttt{if x < y: yield (x, y)}
\end{align*}

\noindent
This nested loop strategy gives both a semantic and operational definition and is exactly the way we also explain
the conceptual evaluation strategy of \SQL in our undergraduate database courses.
In the following, we use this formalism of comprehension of sets (and more generally collections) yet also deviate in three details:
1) We use a tuple instead of a domain perspective.
2) We have an explicit notation of scoping: 
The body can be a logical statement instead of a conjunction of properties, thus the order of shown predicates does not matter. What matters are the well-defined scopes.
3) We use our stricter rule on heads (heads need to be kept clean).

Notice that allowing nesting in the head is usually \emph{the way} 
of defining calculations with collections and list comprehensions.
For example, in Haskell creating all the squares of even numbers is canonically written 
with squaring of numbers happening in the head:
\begin{align*}
&\texttt{[x*x | x<-ns, mod x 2 == 0]}
\end{align*}
Nesting in the head is also useful for the nested relational model, 
and used in extensions of \Datalog.
But nesting in the head is not needed (\cref{sec:head aggregates}) 
and we believe it is distracting 
for the flat (unnested) relational model.

\subsection{Composability through orthogonal nesting}

We allow arbitrary nesting of comprehensions in the body, i.e.\
nesting is orthogonal (and therefore compositional): 
it does not interfere with or restrict other constructs (subject to scoping rules).
For example, in Haskell, the following comprehension is allowed:
\begin{align*}
\texttt{[(x, z) | x <- xs, z <- [ y | y <- ys, x < y]]}
\end{align*}
This expression would correspond in \SQL to the lateral join shown in 
\cref{fig:SQL lateral}.
It is written in \ARC\ as follows:
\begin{align}
  \{ Q(A,B) \mid 
  \, & \exists x \in X, z\in 
  \{ Z(B) \mid \exists y \in Y [Z.B = y.A \wedge x.A < y.A ]\} 
  \label{eq: lateral join gtrc}\\
    & [Q.A = x.A \wedge Q.B = z.B] \}     \notag
\end{align}

\begin{figure}[t]
  \centering
  \begin{subfigure}[b]{.45\linewidth}
  \centering
\begin{minipage}{0.9\linewidth}
\begin{lstlisting}
select x.A, z.B
from X as x
join lateral (
  select y.A as B
  from Y as y
  where x.A < y.A) as z 
on true
\end{lstlisting}
\end{minipage}
\vspace{-7mm}
\caption{}
\label{fig:SQL lateral}
  \end{subfigure}
\hspace{1mm}
\caption{(a): Nested \ARC\ from \cref{eq: lateral join gtrc} expressed as lateral join in \SQL.}
\end{figure}

\subsection{Grouping and aggregates in set semantics}
\label{sec: grouping}

Consider the task of summing the salaries of all employees in a table. 
Under set semantics, projecting the salary column \emph{before applying
an aggregate function removes duplicates}, so we obtain the sum of
distinct salaries rather than the sum over all employees. 
This is usually
not the intended behavior when computing aggregates.

A conceptually simple fix is to apply aggregate functions 
not over individual columns but over entire sets of tuples, 
with each aggregate operating on a designated position of those tuples (e.g., the last position in the unnamed perspective). 
This is the approach proposed by Klug~\cite{DBLP:journals/jacm/Klug82} 
in his classical 1982 paper, 
in which each aggregate function receives \emph{its own separate scope}.\footnote{Like \Datalog, Klug works in the unnamed (positional) perspective. 
To treat aggregates as ordinary unary function symbols on relations, each aggregate must be tied to a fixed column in advance.
Consequently, applying a sum to the 2nd or the 3rd column of the same relation requires two \emph{different} aggregate functions (the column index is effectively baked into the function name). 
This need for separate aggregates per column complicates the formalism notably.}
Subsequent comprehension-based database programming languages (DBPLs)~\cite{DBLP:conf/dbpl/Trinder91,DBLP:journals/jiis/GrustS99,DBLP:journals/sigmod/BunemanLSTW94, DBLP:journals/tods/FegarasM00},
including extensions of logic with aggregate operators~\cite{DBLP:journals/jacm/HellaLNW01}, 
and modern \Datalog-inspired systems such as Souffl\'e~\cite{souffle}
and Rel~\cite{DBLP:conf/sigmod/ArefGKLMMMMNPRS25}, 
inherited variants of this formalism.

This formalism has left two important legacies for set-based languages:
(1) 
Evaluating two different aggregate functions over the same relation 
requires two logical copies of that relation because each aggregate is evaluated in its own independent scope.
(2) Except for Rel~\cite{DBLP:conf/sigmod/ArefGKLMMMMNPRS25}, 
aggregate functions in these formalisms return only their computed results, not the grouped attributes. 
These attributes must instead be specified \emph{outside} the aggregation scope and passed into it via correlated nesting
(a pattern we refer to as ``from the outside in'' = \HL{FOI}). 

We discuss these issues in detail after introducing \ARC's handling of aggregates.
Notice how \ARC\ serves as a \emph{reference language that provides the vocabulary and structure} needed to compare and reason about the behavior of different languages. 
In doing so, \ARC\ \emph{abstracts from the idiosyncrasies of each language's surface syntax and instead captures their shared underlying semantic structure}.

\introparagraph{Aggregation in \ARC}
\ARC's collection framework supports a conceptual evaluation strategy in which \emph{aggregates are defined over the full join}:\footnote{Recall that a full join of two or more relations contains no duplicates (i.e.\ it is a set) if the input relations are sets as well.} 
values are accumulated one element at a time, and multiple aggregates can be evaluated in parallel by reusing the same scope, as in SQL. 
Thus, conceptually, an aggregate has two inputs: the full join (determined by the scope 
in which the \HL{aggregation predicate} appears) and a column identifier
(i.e.\ range variables and column name, as in \TRC).
If desired, deduplication of the input values can be expressed either by first applying a projection 
or by using dedicated aggregate functions 
(e.g., \sql{countdistinct} instead of \sql{count}).
To enforce set semantics of the output, 
deduplication is applied to the result values of the scope.
This is a flexible pattern that allows \ARC\ to recover the different interpretations of aggregates from existing relational languages.

Recall that our aim is not to retrofit aggregates into classical first-order logic~\cite{DBLP:journals/jacm/HellaLNW01}, 
but to provide an abstract calculus that can model the diverse computational patterns 
(including aggregation) found in real relational languages. 
\HA{Relational Calculus}[relational calculus] here is meant as a more general term than first-order logic.
As Date writes~\cite[Ch.~8]{date2004introduction} 
``\emph{calculus ... provides a notation for stating the definition of that
desired relation in terms of those given relations}'', 
and ``\emph{A fundamental feature of the calculus is the range variable}'',
and ``\emph{variable that ``ranges over'' some specified relation}.''
This definition of calculus naturally motivates our proposal of Abstract Relational Calculus (\ARC), an abstract relational query language defined in a collection framework that strictly generalizes \TRC.

Consider a simple grouped aggregate query over a binary relation $R(A,B)$
that computes, for each distinct value of $R.A$, the sum of associated $R.B$ values.
\ARC\ expresses this query in comprehension syntax as follows:
\begin{align}
  \{&Q(A, sm) \mid  
  \exists r \in R, \gamma_{r.A} 
  [Q.A=r.A
  \wedge Q.\textit{sm}=\sql{sum}(r.B) ] \}  
  \label{ALT:simple grouped by aggregate}
\end{align}
The \emph{\HA{aggregation predicate}} $Q.\textit{sm}=\sql{sum}(r.B)$ 
accumulates and aggregates values over the set of tuples produced by the conceptual evaluation strategy.\footnote{Notice that the aggregation predicate serves here simultaneously also as \HL{assignment predicate},
although aggregation predicates can also serve as comparison predicates as we see later.}
Thus, aggregates appear as operands in predicates. 
The query has a \emph{\HA{grouping operator}} $\gamma$ with \emph{grouping key} $r.A$ that partitions 
the full join result within the quantifier scope into groups based on bindings of the grouping key $r.A$.
When the aggregate is taken over the entire join result, then we write $\gamma_\emptyset$ explicitly 
(similar to ``group by true'' in \SQL). 
Thus, the appearance of any aggregation predicate turns an existential scope into a grouping scope
and requires a grouping operator.

\begin{figure}[t]
  \centering   
  \begin{subfigure}[b]{0.99\linewidth}
  \centering		
\begin{minipage}{0.41\linewidth}
\begin{lstlisting}
select R.A, sum(R.B) sm
from R 
group by R.A
\end{lstlisting}
\end{minipage}
\vspace{-7mm}
\caption{Grouped aggregate in \SQL}
\label{fig:SQL:simple_aggregate}
  \end{subfigure}    

  \begin{subfigure}[b]{1.0\linewidth}
    \centering		
    \includegraphics[scale=0.5]{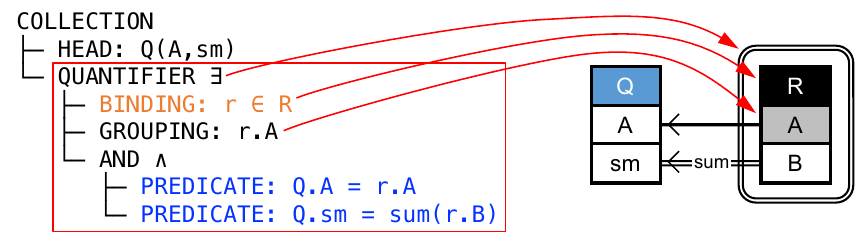}
    \vspace{-1mm}
    \caption{Grouped aggregate in \ARC\ as ALT and higraph modalities}
    \label{Fig_simple_aggregate_ARC}
  \end{subfigure}    

  \caption{The semantics of a simple grouped aggregate query in a ``\emph{\HL{from the inside out}[FIO]}'' pattern represented in \SQL (a) and \ARC\ (b),
  \cref{ALT:simple grouped by aggregate}.
  Red overlay arrows indicate scoping, binding and grouping.}
  \vspace{-2mm}
  \label{Fig_simple_aggregate}
\end{figure}

In the comprehension syntax and ALT, the grouping operator is a child of the quantification scope and turns this scope into a \emph{\HA{grouping scope}}.
In the higraph modality, grouped attributes are highlighted with a gray shade, and the scope is drawn with a double-lined boundary to indicate a grouping scope
(\cref{Fig_simple_aggregate_ARC}).

\introparagraph{From the inside out (\HA{FIO})}
In \ARC, grouping and aggregation happen on attributes inside a scope, 
and the resulting grouped and aggregated attributes are then available outside that  scope. 
We therefore call this pattern ``\emph{from the inside out}.''
It corresponds exactly to the way \SQL\ would represent the grouped aggregate query 
(\cref{fig:SQL:simple_aggregate})
\cite[Database 720]{7240:activities},
and to extended relational algebra:
\begin{align*}
  &\gamma_{A, \textrm{sum}(B)\rightarrow \textrm{sm}} [R]
\end{align*}
Rel~\cite{DBLP:conf/sigmod/ArefGKLMMMMNPRS25} has the same relational pattern:
\begin{align}
&\texttt{def Q(a,sm) : sm = sum[(b) : R(a,b)] } \notag
\end{align}
Rel's interpretation of aggregate queries is that of variable elimination~\cite{DBLP:conf/pods/KhamisNR16}, 
and the query as written in Rel can be viewed as creating a lookup function $f$ with
$f(a) := \sum_{b:R(a,b)} b$.

\introparagraph{From the outside in (\HA{FOI})}
Several languages represent our simple grouped aggregate query in a ``per-outer-tuple'' pattern.
Klug's formalism~\cite{DBLP:journals/jacm/Klug82} uses an unnamed variant of a tuple relational calculus that uses nesting in the head. 
Slightly changing the syntax and porting it to a named perspective, 
the query would be written as follows:\footnote{Original notation:
$(v_{1}[1], \mathrm{sum}_{2} (
    (v_{2}[1], v_{2}[2]) : R(v_{2}) : v_{2}[1] = v_{1}[1]
  )
)
: R(v_{1})$.
}
\begin{align}
\{(r.A, \mathrm{sum}_{2}
  \{ (r_2.A, r_2.B) \mid r_2 \in R \wedge r_2.A = r.A \})
\mid r \in R \} \label{Klug: simple}
\end{align}
Here, the first range variable $r \in R$ fixes the grouping key $r.A$, 
and the second range variable $r_2 \in R$ performs the aggregation for each value
in $r.A$.

Hella et al.~\cite{DBLP:journals/jacm/HellaLNW01}
use the same pattern, but in an unnamed domain perspective:
\begin{align}
  (\exists y. R (x, y))  \wedge (q = \sql{Aggr}_{\Sigma}  \; z. (R (x, z), z))
  \label{eq:libkin formalism:easy}           
\end{align}
In this query, $x$ is a free variable,
and the conjunct $\exists y. R (x, y)$ 
range-restricts its admissible values to those that occur in $R.A$.
The aggregate 
$\sql{Aggr}_{\Sigma}  \; z. (R (x, z), z)$
binds $z$ and is parameterized by the free variable $x$, 
which serves as the grouping key.

In Souffl\'e~\cite{souffle,DBLP:conf/cc/ScholzJSW16}, the query
follows the same pattern:
\begin{align}
&\texttt{Q(a,sum b: \{R(a,b)\}) :- R(a,\_).} \label{souffle:head aggregate}
\end{align}
The documentation~\cite{souffle} describes this pattern explicitly: ``\emph{You cannot export information from within the body of an aggregate. This means that you cannot ground a variable from within the scope of the aggregate body and expect this grounding to transfer to the outer scope}.''

\begin{figure}[t]
  \centering   
  \begin{subfigure}[b]{.39\linewidth}
  \centering		
\begin{minipage}{1\linewidth}
\begin{lstlisting}
select distinct R.A,
  (select sum(R2.B) sm
  from R R2
  where R2.A=R.A)
from R
\end{lstlisting}
\end{minipage}
\vspace{-7mm}
\caption{Scalar subquery in \SQL}
\label{fig:SQL_souffle_scalar}
  \end{subfigure} 
\hspace{3mm}
  \begin{subfigure}[b]{.44\linewidth}
  \centering		
\begin{minipage}{1\linewidth}
\begin{lstlisting}
select distinct R.A, X.sm
from R join lateral
  (select sum(R2.B) sm
  from R R2
  where R2.A=R.A) X
on true
\end{lstlisting}
\end{minipage}
\vspace{-7mm}
\caption{Lateral join in \SQL}
\label{fig:SQL_souffle_lateral}
  \end{subfigure}    

\begin{subfigure}[b]{1.0\linewidth}
  \centering		
  \vspace{1mm}
  \includegraphics[scale=0.5]{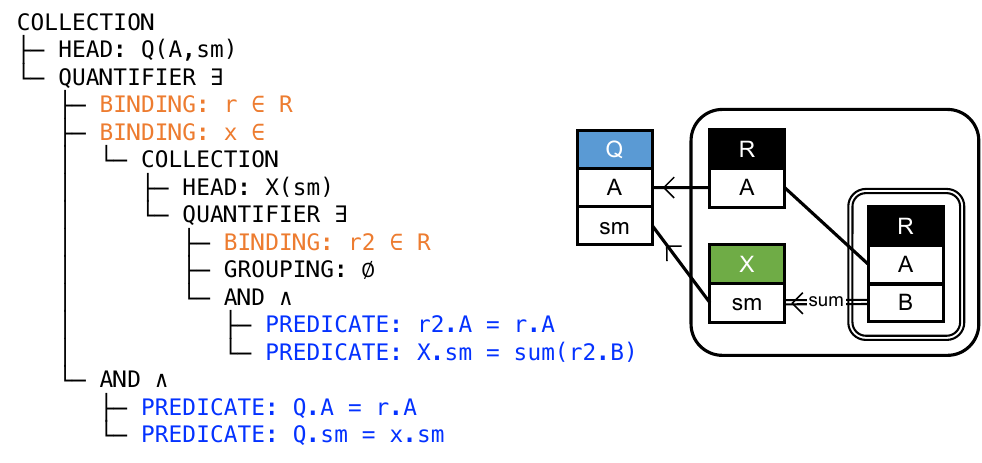}
  \vspace{-3mm}
  \caption{FOI pattern in \ARC}
  \label{Fig_souffle_ARC}
\end{subfigure}    

\caption{The semantics of a simple grouped aggregate query in a ``\emph{\HL{from the outside in}[FOI]}'' pattern represented in \SQL with a scalar subquery (a) or lateral join (b), 
and \ARC\ (c),
\cref{ARC: FOI comprehension}.
This relational pattern corresponds to the way Klug~\cite{DBLP:journals/jacm/Klug82}
\cref{Klug: simple},
Hella et al.~\cite{DBLP:journals/jacm/HellaLNW01} \cref{eq:libkin formalism:easy},
and
Souffl\'e~\cite{souffle,DBLP:conf/cc/ScholzJSW16} \cref{souffle:head aggregate} express the query.
}
\label{Fig_FOI_pattern}
\end{figure}

This pattern of using two range variables over the same relation
to perform a grouped aggregate query without an explicit \sql{GROUP BY} clause
can be represented in \SQL either via scalar subqueries or via lateral joins.
\Cref{fig:SQL_souffle_scalar} and \cref{fig:SQL_souffle_lateral} show two different syntactic variants in SQL
that represent this pattern. 
Notice that both queries are semantically equivalent: 
a \HL{single-valued scalar query}[single-valued] can always be written as a lateral join (see \cref{sec:head aggregates}).

In \ARC, the same pattern expressed as ALT, higraph, and comprehension syntax is shown in \cref{Fig_souffle_ARC} and as:
\begin{align}
\{Q&(A, sm) \mid \exists r \in R, x \in \{X(sm) \mid  \exists r_2 \in R, \gamma_{\emptyset} 
  [r2.A=r.A \, \wedge          \label{ARC: FOI comprehension} \\
& X.\sql{sm}=\sql{sum}(r_2.B)] \} 
[Q.A=r.A \wedge Q.\sql{sm}=x.\sql{sm}]\}       \notag 
\end{align}
Notice that while \SQL does not use an explicit grouping clause 
when an aggregate is computed over the entire relation,
\ARC\ indicates the aggregate evaluation explicitly with a grouping on the empty set: there is just one group, an aggregate is evaluated over all tuples 
(similar to ``\sql{group by true}'' in \SQL), and \ARC\ makes this explicit.

Notice that 
\ARC\ introduces an explicit intermediate 
\HL{defined relation} $X$ 
that exists only implicitly in the surface syntax of several languages.
While function composition, 
as in $(g \circ f)(x) = g(f(x))$ 
or the head-nested scalar subquery, hides an intermediate relation, the lateral-join formulation in \SQL\ already exposes it as a derived relation.
Recall that an abstract relational query language serves as a \emph{reference language} and 
\emph{should make implicit patterns explicit}.
Thus, \ARC\ represents these conceptual structures explicitly as defined relations,
which provides a clear abstraction for understanding how queries are built in a modular way from smaller components,
even when they are never or cannot be materialized (see \cref{sec:defined relations}).
Also notice that we do not need to name them in the higraph modality (\cref{Fig_souffle_ARC}):
they exist on the Canvas as independent topological entities and may remain unnamed.

\begin{figure}[t]
  \centering   
  \begin{subfigure}[b]{.49\linewidth}
  \centering		
\begin{minipage}{1\linewidth}
\begin{lstlisting}
select R.dept, avg(S.sal) av
from R, S 
where R.empl=S.empl
group by R.dept
having sum(S.sal)>100
\end{lstlisting}
\end{minipage}
\vspace{-7mm}
\caption{Multiple aggregates in \SQL}
\label{fig:SQL_Hella2001}
  \end{subfigure}    

  \begin{subfigure}[b]{.8\linewidth}
    \centering		
    \includegraphics[scale=0.5]{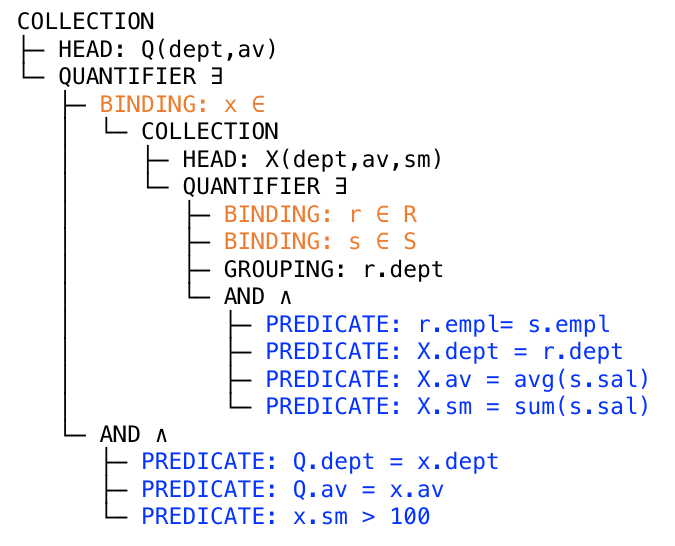}
    \vspace{-1mm}
    \caption{\ARC\ ALT modality}
    \label{fig:Fig_Libkin2001_2}
  \end{subfigure}    

  \begin{subfigure}[b]{.48\linewidth}
    \includegraphics[scale=0.5]{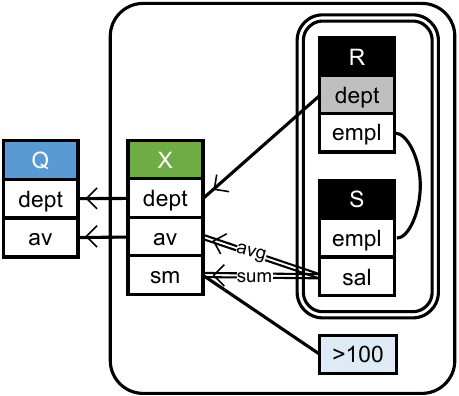}
    \vspace{-1mm}
    \caption{\ARC\ higraph modality}
  \end{subfigure}
    \caption{
    (a): Running example for multiple aggregates from \cite{DBLP:journals/jacm/HellaLNW01} in \SQL (a) and \ARC\ (b), (c), \cref{arc: hella simple}.}
    \vspace{-2mm}
    \label{Fig_Libkin2001_1}
\end{figure}

\introparagraph{Multiple aggregates}
We illustrate our formalism using the running example from Hella et al.~\cite{DBLP:journals/jacm/HellaLNW01} 
``\emph{returning the average salary for each department that pays total salary at least $100$}''
over a schema 
(with slightly simplified relation names and constant)
$R(\textit{empl}, \textit{dept}), S(\textit{empl},\textit{sal})$ 
representing employees, their departments, and their salaries.
\Cref{fig:SQL_Hella2001} shows the corresponding \SQL query 
\cite[Database 740]{7240:activities}.

In \ARC, a \sql{HAVING} clause is simply a selection applied after an aggregation:
\begin{align}
  \{&Q(\textit{dept},\textit{av}) \mid  
  \exists x \in \{X(\textit{dept}, \textit{av}, \textit{sm}) \mid 
  \exists r \in R, s \in S,
  \gamma_{r.\textit{dept}}           \label{arc: hella simple}\\[-1mm]
  &[X.\textit{dept}=r.\textit{dept}
  \wedge X.\textit{av}=\sql{avg}(s.\textit{sal}) 
  \wedge X.\textit{sm}=\sql{sum}(s.\textit{sal}) \, \wedge   \notag \\[-1mm]
  &r.\textit{empl} = s.\textit{empl}] \} 
  [Q.\textit{dept} = x.\textit{dept}
  \wedge Q.\textit{av}=x.\textit{av}
  \wedge x.\textit{sm}>100] \}     \notag
\end{align}

\begin{figure}[t]
    \centering        
    \includegraphics[scale=0.5]{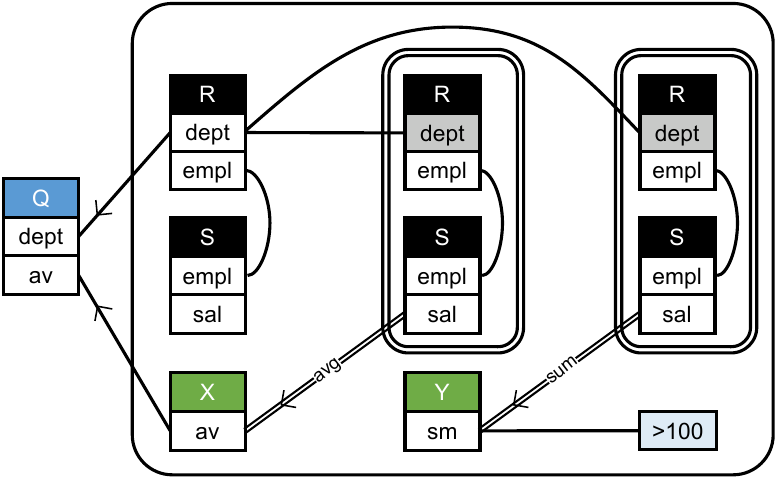}
    \caption{Pattern-preserving \ARC\ higraph representation of the multiple-aggregate query
    in the formalism by Hella et al.~\cite{DBLP:journals/jacm/HellaLNW01} 
    \cref{eq:libkin formalism}.}
    \vspace{-1mm}
    \label{Fig_Libkin2001_2}
\end{figure}

The query expressed in the language 
$\mathcal{L}_{\textrm{aggr}} ( \{ < \} , \{ \sum , \sql{AVG} \} )$
by Hella et al.~\cite{DBLP:journals/jacm/HellaLNW01} 
defines an output relation
$Q(y,q)$ via the following expression:\footnote{We made an adjustment to the query after confirming with the authors that our interpretation was correct. The aggregation as originally written in \cite{DBLP:journals/jacm/HellaLNW01},
$\sql{Aggr}_{\sum} z. (\exists y.R (x, y) \wedge S (x, z), z)$ 
would, on a database containing two employees with the same salary in the same department, count that salary only once rather than twice. Also, ``$c_{100}$'' is a language-specific syntax for referring to the constant 100.}
\begin{align}
  Q(y,q) := \;
  &(\exists x \exists z. R (x, y) \wedge  S (x, z))  \label{eq:libkin formalism}           \\[-1mm]
  &\wedge (q = \sql{Aggr}_{\textrm{AVG}} x,z. (R (x, y) \wedge S (x, z), z))   \notag \\[-1mm]
  &\wedge (\sql{Aggr}_{\sum} x,z. (R (x, y) \wedge S (x, z), z) > c_{100} )  \notag
\end{align}
For a fixed $y$, the aggregate term
$\operatorname{Aggr}_{\Sigma} x,z. (R(x,y) \wedge S(x,z),z)$
ranges over all \emph{distinct} rows $(x,z)$ such that $R(x,y) \wedge S(x,z)$ holds. It collects the bag
$\dgal{z \mid \exists x, z [R(x,y) \wedge S(x,z)]}$ 
(with multiplicities)
and applies the summation operator
$\Sigma$ (sum) to that bag.
This formalism (inherited from Klug~\cite{DBLP:journals/jacm/Klug82}) changes the signature of the query:
the same base relations are referenced multiple times, 
once in each aggregation scope,
and once outside the aggregation scopes.
This leads to a \emph{modified relational pattern}, 
shown in the higraph modality (\cref{Fig_Libkin2001_2}) and in comprehension syntax modalities:
\begin{align}
  \{&Q(\textit{dept},\textit{av}) \mid  
  \exists r_3 \in R, s_3 \in S,                             \label{ARC:Libkin:2001} \\[-1mm]
  &x \in \{X(\textit{av}) \mid 
  \exists r_1 \in R, s_1 \in S, \gamma_{r_1.\textit{dept}}       \notag  \\[-1mm]
  &[r_1.\textit{dept}=r_3.\textit{dept}
  \wedge r_1.\textit{empl}=s_1.\textit{empl} 
  \wedge X.\textit{av}=\sql{avg}(s_1.\textit{sal}) ] \},   \notag \\[-1mm]
  &y \in \{Y(\textit{sm}) \mid 
  \exists r_2 \in R, s_2 \in S, \gamma_{r_2.\textit{dept}}    \notag \\[-1mm]
  &[r_2.\textit{dept}=r_3.\textit{dept}
  \wedge r_2.\textit{empl}=s_2.\textit{empl} 
  \wedge Y.\textit{sm}=\sql{sum}(s_2.\textit{sal}) ] \}    \notag \\[-1mm]
  &[Q.\textit{dept} = r_3.\textit{dept}
  \wedge Q.\textit{av}=x.\textit{av}
  \wedge r_3.\textit{empl}=s_3.\textit{empl}
  \wedge y.\textit{sm}>100 ] \}     \notag
\end{align}

\begin{figure}[t]
    \centering        
    \includegraphics[scale=0.5]{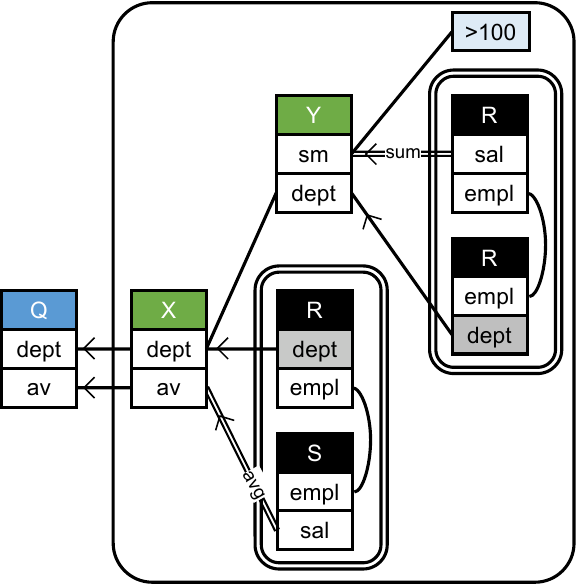}
    \caption{Pattern-preserving \ARC\ higraph representation of the multiple-aggregate query written in Rel~\cref{rel: Hella aggregates}.}
    \vspace{-1mm}
    \label{Fig_Hella_Rel_pattern}
\end{figure}

While Rel~\cite{rel_cheatsheet,DBLP:conf/sigmod/ArefGKLMMMMNPRS25} follows the \HL{FOI} pattern for aggregation, 
it still inherits the pattern of using distinct aggregation scopes (i.e.\ separate subqueries) for each aggregate over the same relation:
\begin{align}
\texttt{def }&\texttt{Q(d,av) :}                      \label{rel: Hella aggregates}       \\[-1mm]
&\texttt{av = average[(e,s) : R(e,d) and S(e,s)] and}       \notag \\[-1mm]
&\hspace{14.2mm}\texttt{sum[(e,s) : R(e,d) and S(e,s)] > 100}      \notag 
\end{align}
\Cref{Fig_Hella_Rel_pattern} shows this relational pattern 
in \ARC's higraph modality, and the corresponding comprehension syntax is:
\begin{align}
  \{&Q(\textit{dept},\textit{av}) \mid    \label{ARC:Rel multiple aggregates} \\[-1mm]
  &x \in \{X(\textit{dept}, \textit{av}) \mid 
  \exists r_1 \in R, s_1 \in S, \gamma_{r_1.\textit{dept}}      \notag   \\[-1mm]
  &[X.\textit{dept}=r_1.\textit{dept}
  \wedge r_1.\textit{empl}=s_1.\textit{empl} 
  \wedge X.\textit{av}=\sql{avg}(s_1.\textit{sal}) ] \},   \notag \\[-1mm]
  &y \in \{Y(\textit{dept}, \textit{sm}) \mid 
  \exists r_2 \in R, s_2 \in S, \gamma_{r_2.\textit{dept}}    \notag \\[-1mm]
  &[Y.\textit{dept}=r_2.\textit{dept}
  \wedge r_2.\textit{empl}=s_2.\textit{empl} 
  \wedge Y.\textit{sm}=\sql{sum}(s_2.\textit{sal}) ] \}    \notag \\[-1mm]
  &[Q.\textit{dept} = x.\textit{dept}
  \wedge Q.\textit{av}=x.\textit{av}
  \wedge x.\textit{dept}=y.\textit{dept}
  \wedge y.\textit{sm}>100 ] \}     \notag
\end{align}
Notice the similarities and differences between the relational patterns
of
\cref{Fig_Hella_Rel_pattern}/\cref{ARC:Rel multiple aggregates}
and 
\cref{Fig_Libkin2001_2}/\cref{ARC:Libkin:2001}.

\begin{figure}[h]
\centering	
\begin{subfigure}[b]{.39\linewidth}		
\begin{lstlisting}
select exists(
  select 1
  from R 
  where R.q <= 
    (select count(S.d)
    from S
    where S.id=R.id))
\end{lstlisting}
\vspace{-6mm}
\caption{}
\label{sql:sentence_with_aggregate1}
\end{subfigure}	
\hspace{6mm}
\begin{subfigure}[b]{.37\linewidth}
    \includegraphics[scale=0.5]{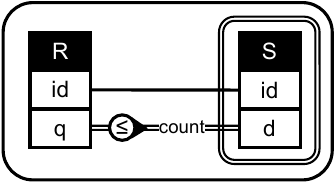}
\vspace{2mm}
\caption{}
\label{Fig_sentence_with_aggregate1}
\end{subfigure}	

\begin{subfigure}[b]{.39\linewidth}		
\begin{lstlisting}
select not exists(
  select 1
  from R 
  where R.q >
    (select count(S.d)
    from S
    where S.id=R.id))
\end{lstlisting}
\vspace{-6mm}
\caption{}
\label{sql:sentence_with_aggregate2}
\end{subfigure}	
\hspace{6mm}
\begin{subfigure}[b]{.37\linewidth}
    \includegraphics[scale=0.5]{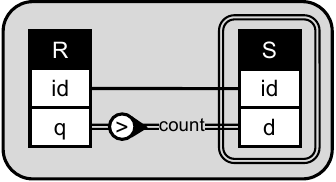}
\vspace{2mm}
\caption{}
\label{Fig_sentence_with_aggregate2}
\end{subfigure}	

\caption{Boolean queries and constraints \cref{arc:statement 1}, \cref{arc:statement 2}.}
\label{Fig_sentence_with_aggregate}
\end{figure}

\introparagraph{Logical sentences and integrity constraints}
Expressions that evaluate to true or false can also contain aggregates. 
Furthermore, \HL{aggregation predicates}[aggregation predicate] may be \HL{comparison predicates}[comparison predicate], not \HL{assignment predicates}[assignment predicate].
\Cref{Fig_sentence_with_aggregate1,Fig_sentence_with_aggregate2}
show two \ARC\ sentences that illustrate this pattern (see \cite{disjunctions} on how to read the negation scope):
\begin{align}
& \exists r \in R[
  \exists s \in S, \gamma_{\emptyset}[
  r.id =s.id \wedge r.q \leq \sql{count}(s.d)]]     \label{arc:statement 1}\\
& \neg \exists r \in R[
  \exists s \in S, \gamma_{\emptyset}[
  r.id =s.id \wedge r.q > \sql{count}(s.d)]]        \label{arc:statement 2}
\end{align}
By contrast, the closest \SQL\ formulations, shown in \Cref{sql:sentence_with_aggregate1,sql:sentence_with_aggregate2}, can only return a unary relation representing the truth value, not a Boolean sentence directly~\cite[Database 737]{7240:activities}.

\subsection{Language \HA{Conventions}[convention]}
\label{sec:conventions}

Consider an instance with $R=\{(1,2)\}$ and $S=\emptyset$.
The following Souffl\'e rule computes, for each $R(ak,\_)$, the sum of all $b$ such that $S(a,b)$ and $a<ak$:
\begin{align}
\texttt{Q(ak,sm) :- R(ak,\_), sm = sum b: \{S(a,b), a<ak\}.}
\label{souffle:sum-empty}
\end{align}
On this instance, the rule derives $Q(1,0)$ because Souffl\'e evaluates a sum over an empty set as $0$ (Souffl\'e has no \texttt{NULL}).
In contrast, the equivalent \SQL queries in \cref{fig:sql scalar query,fig:sql lateral query} (if we add  DISTINCT to the select clause)
return the row $(1,\texttt{NULL})$ on the same instance, since in SQL the result of \texttt{SUM} over zero input rows is \texttt{NULL}.

We treat such choices
(how aggregates behave on empty inputs, 
and more generally how missing values are represented) as \emph{conventions}.
They are orthogonal to the relational structure of the query (see \cref{Fig_scalar_query_separation}): changing the convention affects the observable result, but not the underlying relational pattern.
Accordingly, \ARC\ abstracts from these conventions and focuses on the relational composition of queries.

\subsection{Sets or bags? Not an issue for \ARC\ but a matter of convention}
\label{sec:bags}

Nothing needs to change \emph{in the surface syntax of \ARC{}} 
if relations are interpreted as bags (multisets) rather than sets.
The conceptual evaluation still ranges over tuples as before: 
each tuple in one relation can be paired with each tuple in the other relation,
regardless of whether a tuple has a duplicate or not.
Consequently, a relational QL does not need to be designed for sets or bags; 
instead the same query can be \emph{interpreted} under either set or bag semantics. 
Choosing set or bag interpretation is orthogonal to language design.

A common convention in the collection-types literature is to signal bag semantics by writing bag brackets (here: $\dgal{\cdot}$) instead of set brackets $\{\cdot\}$, e.g.,
$\dgal{Q(A) \mid r \in R[Q.A =r.A]}$
instead of
$\{Q(A) \mid r \in R[Q.A =r.A]\}$.
However, we treat this as a convention rather than part of the concrete syntax of query strings.
The syntax of \ARC{} does not commit to sets or bags, 
and the choice of semantics is fixed independently of the relational patterns expressed by the query.

Whether a query is interpreted under set or bag semantics matters for
evaluation and optimization, 
because some rewrite rules only apply under set semantics.
For example, consider the nested query
\begin{align*}
\{Q(A) \mid \exists r \in R [\exists s \in S 
[Q.A = r.A \wedge r.B=s.B]]\}
\end{align*}
Under set semantics this can be unnested to
\begin{align*}
\{Q(A) \mid \exists r \in R, s \in S 
[Q.A = r.A \wedge r.B=s.B]\}
\end{align*}
Under bag semantics, however, the two can differ: the nested formulation produces $Q(A)$ once per matching occurrence of $r$ (a semijoin-like behavior),
whereas the unnested formulation produces $Q(A)$ once per matching pair $(r,s)$, which multiplies output multiplicities when multiple tuples of $S$ share the same $B$-value.

\introparagraph{Deduplication}
Removing duplicates (as in \sql{DISTINCT}) is expressible via grouping on all projected attributes and does not require a dedicated operator.
For example, deduplicating a binary relation $R(A,B)$ can be written as:
\[
  \{Q(A,B) \mid \exists r \in R,\ \gamma_{r.A, r.B}[\,Q.A = r.A \wedge Q.B = r.B\,]\}.
\]
Recall that an aggregate predicate entails a grouping clause, 
but grouping can also appear without having an aggregate predicate.

\begin{figure}[t]
  \centering
  \begin{subfigure}[b]{.43\linewidth}
  \centering
      \includegraphics[scale=0.45]{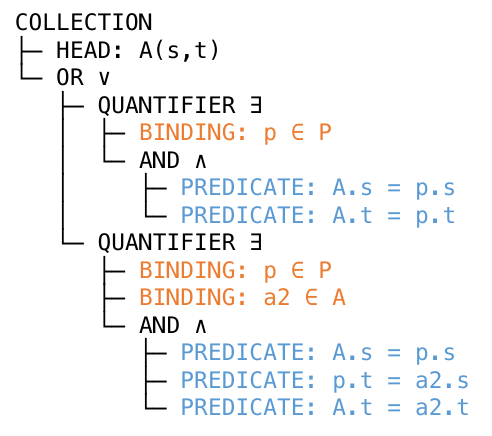}
      \vspace{-2mm}
      \caption{}
      \label{Fig_Recursion_ALT}
  \end{subfigure}
	\hspace{1mm}
  \begin{subfigure}[b]{.52\linewidth}
  \centering		
      \vspace{1mm}		
      \includegraphics[scale=0.5]{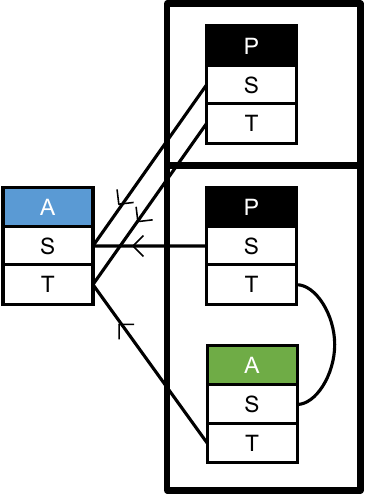}
      \vspace{-1mm}
      \caption{}
      \label{Fig_Recursion_RD}
  \end{subfigure}
\caption{\ARC\ representations for recursive query \cref{eq:GTRC recursion}.}
\end{figure}

\subsection{Negation, Disjunction, and Union}

Union of relations is treated as disjunction in \TRC{} and \ARC.
Negation and disjunction are discussed in detail in~\cite{disjunctions}.

\subsection{Recursion}
\label{sec:recursion}

\ARC\ supports recursion with the same least-fixed-point semantics as \Datalog, but expressed in our \emph{named} perspective.
Let $P(s,t)$ be the parent relation, where $s$ is the source (parent) and $t$ is the target (child).
In \Datalog, the ancestor relation $A(s,t)$ is defined by the familiar two-rule program:
\begin{align}
&\texttt{A(x,y) :- P(x,y)}                      \notag\\[-1mm]
&\texttt{A(x,y) :- P(x,z), A(z,y)}              \notag
\end{align}

In \Datalog, multiple rules with the same head are combined by union, 
and recursion is obtained by taking the least fixed point of that union.
In \ARC, a relation is defined by a single construct
and the implicit union of multiple rules is written as a disjunction within one definition.
\begin{align}
\{A(s,t) \mid \,
& \exists p \in P 
  [A.s=p.s
  \wedge A.t =p.t]
  \,\vee    \label{eq:GTRC recursion}\\
& \exists p \in P, a_2 \in A
  [A.s=p.s
  \wedge p.t =a_2.s
  \wedge a_2.t =A.t] \} \notag
\end{align}

\subsection{Null values and (NOT) IN predicates}

\begin{figure}[h]
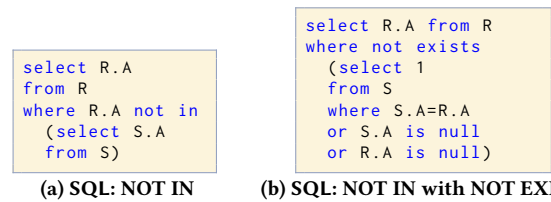

  \centering
  \begin{subfigure}[b]{.43\linewidth}
  \centering
\begin{minipage}{0.68\linewidth}
\begin{lstlisting}
select R.A 
from R 
where R.A not in
  (select S.A 
  from S)
\end{lstlisting}
\end{minipage}
\vspace{-7mm}
\caption{\SQL: NOT IN}
\label{fig:SQL_NOT_IN}
  \end{subfigure}
  \begin{subfigure}[b]{.52\linewidth}
  \centering		
\begin{minipage}{0.72\linewidth}
\begin{lstlisting}
select R.A from R 
where not exists
  (select 1
  from S
  where S.A=R.A
  or S.A is null
  or R.A is null)
\end{lstlisting}
\end{minipage}
\vspace{-7mm}
\caption{\SQL: NOT IN with NOT EXISTS}
\label{fig:SQL_NOT_IN_with_EXISTS}
  \end{subfigure}
\caption{Replicating SQL's null behavior for the NOT IN clause (a) with NOT EXISTS (b).
\Cref{eq: null GTRC} shows their \ARC{} representation. 
}
\label{fig:SQL_NOT_IN_and_EXISTS}
\end{figure}

SQL evaluates predicates in three-valued logic, so comparisons involving null
may yield unknown. This interacts poorly with certain predicates, notably
\sql{NOT IN}. For example, the \SQL query in \cref{fig:SQL_NOT_IN} returns the empty set
whenever $S$ contains any row with null in column $A$, because the membership test
becomes unknown and the \sql{WHERE} clause filters out the row.

But this behavior can be reproduced
within two-valued logic by rewriting \sql{NOT IN} into \sql{NOT EXISTS} while making
null checks explicit~\cite{DBLP:conf/pods/LibkinP23} as in 
\cref{fig:SQL_NOT_IN_and_EXISTS}.
We can thus replicate SQL's \texttt{NULL} behavior in our collection framework as well:
\begin{align}
\{Q(A) \mid \,
& \exists r \in R 
  [Q.A=r.A \,
  \wedge    \label{eq: null GTRC}\\
& \neg (\exists s \in S [s.A=r.A
  \vee s.A \sql{ is null}
  \vee r.A \sql{ is null} 
  ] ) ] \} \notag
\end{align}

\subsection{Left and full outer joins}

A priori, outer joins are not naturally expressible with plain comprehensions:
comprehensions range over existing collections,
so a binding with no match simply disappears.
For example, a left join between $R$ and $S$ can be written as the union of the matching and the non-matching cases:
\begin{align*}
&\{ Q(A, B) \mid \exists r \in R, s \in S 
  [Q.A \!=\! r.A \wedge Q.B \!=\! s.B \wedge r.A \!=\!s.B ]\}  \, \cup \\
& \{ Q(A, B) \mid \exists r \in R [Q.A \!=\! r.A \wedge Q.B \!=\! \sql{null} 
  \wedge \neg (\exists s \in S 
  [r.A\!=\!s.B] )  ] \}  \notag
\end{align*}

We therefore extend comprehensions with an explicit join annotation in the binding list
(similar in spirit to the grouping operator).
A join annotation specifies 
($i$) which bound tables are combined by inner/left/full joins and
($ii$) the precedence (nesting) of these joins.
With this extension, the left join above can be expressed as the single comprehension
\begin{align*}
\{ Q(A, B) \mid \, 
  & \exists r \in R, s \in S, \sql{left}(r,s) \\
  &[Q.A \!=\! r.A \wedge Q.B \!=\! s.B \wedge r.A\!=\!s.B ]\} 
\end{align*}

The annotation $\sql{inner}$ is $k$-ary, while $\sql{left}$ and $\sql{full}$ are binary.
Any scope without an explicit outer-join annotation is inner by default.
For example,
$\exists r\in R, s\in S, t\in T[\ldots]$
is shorthand for
$\exists r\in R, s\in S, t\in T, \sql{inner}(r,s,t)[\ldots]$,
and an inner join followed by a left join can be written as
$\exists r\in R, s\in S, t\in T, \sql{left}(r,\sql{inner}(s,t))[\ldots]$.
Our join annotations can model arbitrary nestings of outer joins, including cases that are awkward
to express in surface \SQL{} syntax~\cite{David:1999kc,colgan:oracle:2023}.

\begin{figure}[t]
  \centering
  \begin{subfigure}[b]{.41\linewidth}
  \centering
\begin{minipage}{1\linewidth}
\begin{lstlisting}
select R.m, S.n
from R 
left outer join S
on (R.h=11 and R.y=S.y)
\end{lstlisting}
\end{minipage}
\vspace{-5mm}
\caption{Complicated outer join condition expressed in \SQL}
\label{fig:complicated outer join}
  \end{subfigure}
  \begin{subfigure}[b]{.52\linewidth}
  \centering		
    \includegraphics[scale=0.5]{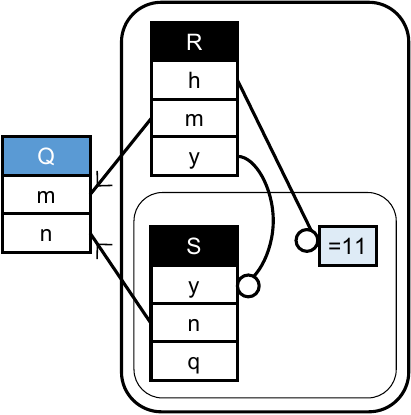}
    \vspace{-1mm}
    \caption{}
    \label{Fig_outer_joins_oracle_N}
  \end{subfigure}
\caption{Outer joins and their higraph representation.}
\end{figure}

At the higraph level, we depict outer join conditions by marking the optional side with an empty circle 
(inspired by ERD notation). Precedence scopes mirror the nesting of join annotations and can also cover cross joins.
For example, \cref{fig:complicated outer join} (from \cite[example~N']{colgan:oracle:2023})
corresponds to:
\begin{align}
\{Q(m,n) \mid \,
& \exists r \in R, s \in S, \sql{left}(r,\sql{inner}(11,s))  \\
&[Q.m=r.m
  \wedge Q.n =s.n
  \wedge r.y=s.y 
  \wedge r.h=11 ] \}    \notag
\end{align}
Here a literal $c$ used as a leaf inside a join annotation denotes a singleton relation
(a virtual unary table) containing just the value $c$; 
hence $\sql{inner}(11,s)$ is a cross join
between $S$ and this singleton.
Because outer cross joins contribute no join-condition edge in the higraph, we annotate them textually
(e.g., with ``$\times$'') when needed.

\begin{figure}[t]
\centering
  \begin{subfigure}[b]{.93\linewidth}
  \centering
  \begin{minipage}{0.4\linewidth}
\begin{lstlisting}
select R.A,
  (select sum(S.B) sm
  from S
  where S.A<R.A)
from R
\end{lstlisting}
  \end{minipage}
  \vspace{-7mm}
  \caption{Scalar query}
  \label{fig:sql scalar query}
  \end{subfigure}
\begin{subfigure}[b]{.37\linewidth}
  \centering
\begin{minipage}{1\linewidth}
\begin{lstlisting}
select R.A, X.sm
from R join lateral
  (select sum(S.B) sm
  from S
  where S.A<R.A) X
on true
\end{lstlisting}
\end{minipage}
\vspace{-7mm}
\caption{Lateral query}
\label{fig:sql lateral query}
\end{subfigure}
  \hspace{2mm}
  \begin{subfigure}[b]{.51\linewidth}
  \centering
\begin{minipage}{0.8\linewidth}
\begin{lstlisting}
select R.A, sum(S.B) sm
from R
left join S
on S.A<R.A
group by R.A
\end{lstlisting}
\end{minipage}
\vspace{-7mm}
\caption{Left join (incorrect translation; 
shown as counterexample)}
\label{fig:sql left join query}
  \end{subfigure}
  \begin{subfigure}[b]{1\linewidth}
  \centering		
    \includegraphics[scale=0.5]{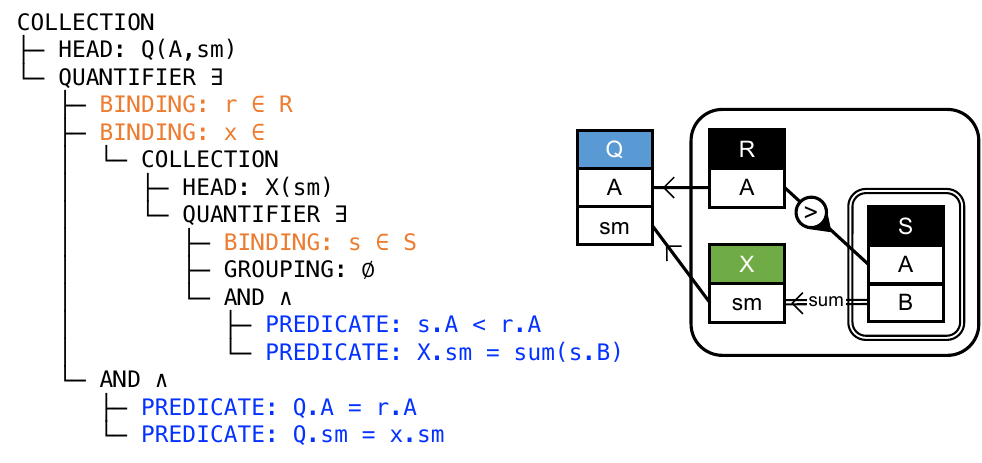}
    \vspace{-3mm}
    \caption{Relational Pattern of a \HL{single-valued} head-nested query}
    \label{Fig_scalar_query_separation}
  \end{subfigure}

\caption{
(a) A \HL{single-valued} correlated scalar SQL query with an aggregate in the head. 
(b) An equivalent formulation that pushes the aggregate into the body using a lateral join. 
(c) An alternative formulation using a left outer join and GROUP BY. Only the lateral-join formulation (b) is guaranteed to preserve the semantics under both set and bag semantics (in particular, when $R$ contains duplicates). 
(d): ARC\ does not allow nesting in the head and therefore represents such scalar queries directly in the lateral-join form (b).}
\label{fig:scalar query rewriting}
\end{figure}

\subsection{Representing head aggregates}
\label{sec:head aggregates}

Recall from 
\cref{sec:set comprehension}
that \ARC\ does not allow nesting (subqueries) in the head,
and from \cref{sec: grouping}
that \ARC\ represents head aggregates as a form of lateral join in the body.
We call a head aggregate in any relational language \emph{\HA{single-valued}} if, 
for every result tuple of the query body, 
the aggregate evaluates to a single scalar value (or null). 
This class includes \sql{SQL} scalar subqueries 
such as \cref{fig:SQL_souffle_scalar},
as well as Soufflé head aggregates such as query \cref{souffle:head aggregate}.
For those queries, the overall result is a flat relation, i.e., it contains no nested collections.
Any single-valued head aggregate can be rewritten as a lateral join in the body.\footnote{This has already been observed in~\cite[Sect.~10]{Bussche:Vansummeren:2009} for set semantics.}
The intuition is that a lateral join faithfully preserves the intended per-tuple semantics of a correlated scalar subquery: 
the inner query is re-evaluated once per outer tuple, without accidental grouping or merging. 
In contrast, a rewrite based on \sql{LEFT JOIN} + \sql{GROUP BY} 
\cite{10.1145/375663.375748, DBLP:journals/tods/FegarasM00}
fails to preserve the correlation pattern under bag semantics 
when grouping coalesces duplicates in the outer relation into a single output row.

For example, consider the \HL{single-valued} \SQL\ scalar subquery in \cref{fig:sql scalar query} and two rewrites:
as \sql{lateral join} in \cref{fig:sql lateral query} and
as \sql{left join} in \cref{fig:sql left join query}~\cite[Database 720]{7240:activities}.
Both rewrites are correct if the inputs contain no duplicates.
Under bag semantics, however, 
if relation $R$ contains duplicate values (and rows don't have a unique key),
then the query in \cref{fig:sql left join query}  collapses all identical $R.A$ values into a single group
and no longer reflects the ``once per tuple of $R$'' evaluation of the subquery.\footnote{If each outer tuple had a unique identifier, then we could add it to the GROUP BY clause and preserve the semantics.}
In contrast, the lateral join in \cref{fig:sql lateral query}
remains equivalent even under bag semantics, 
because the lateral join preserves the per-outer-tuple semantics.
In other words, a \HL{single-valued} scalar subquery and its lateral-join encoding use the same conceptual evaluation strategy. For that reason, \ARC\ 
represents scalar queries as lateral joins (\cref{Fig_scalar_query_separation}).

\subsection{Defined relations (incl.~abstract relations)}
\label{sec:defined relations}

\begin{figure}[t]
\centering
\includegraphics[scale=0.55]{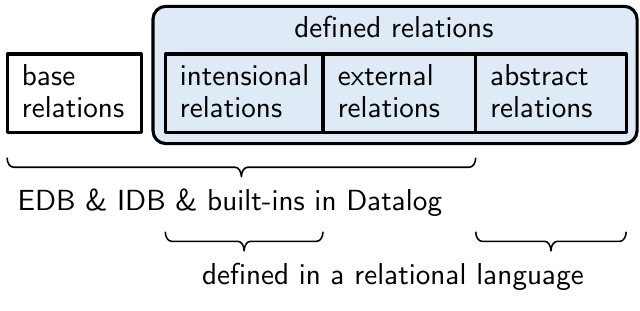}
\caption{\emph{\HL{Base relations}[base relation]} are given extensionally by enumeration.
\emph{\HL{Defined relations}[defined relation]} are given intensionally by definitions.
\emph{\HL{Intensional relations}[intensional relation]} (views, CTEs, IDBs) are defined by relational queries and may be materialized.
\emph{\HL{External relations}[external relation]} (built-ins) 
are defined outside the relational language and may have infinite extension.
\emph{\HL{Abstract relations}[abstract relation]} are possibly domain-dependent relational expressions that 
help abstract and modularize large queries.}
\label{Fig_Table_types}
\end{figure}

In principle, relational query languages can treat functions and arithmetic predicates uniformly as relations.
Unlike \emph{\HA{base relations}[base relation]} (base tables), 
\emph{\HA{defined relations}[defined relation]} are not specified extensionally by enumerating their tuples, 
but intensionally via a definition.
Among defined relations, \emph{\HA{intensional relations}[intensional relation]} (e.g., views and CTEs = Common Table Expressions) 
are definable in the relational language
and, over a finite database instance, have a finite extension (and thus can be materialized).
In Datalog terminology, base and intensional relations correspond to extensional and intensional predicates (EDB and IDB), respectively (\cref{Fig_Table_types}).

\subsubsection{\HA{External relations}[external relation]}
In contrast to intensional relations, external relations (often referred to as external predicates)
are defined outside the relational language and may have infinite extension.
Intuitively, they correspond to built-in predicates (or built-ins) in Datalog, 
i.e.\ predefined relations that extend pure logical atoms with computational or domain-specific functionality, 
e.g.\ the arithmetic predicate ``$+$'', 
equality ``$=$'', 
comparisons such as ``$>$'', 
or string comparison such as SQL's ``\texttt{LIKE}'' operator.\footnote{While 
the comparison operator ``$>$'' may be part of a relational vocabulary, we cannot define a binary relation $\sql{Bigger}(A,B)$
containing pairs of integers where $A>
B$ with relational operations alone.}

\begin{figure}[t]
\centering
\begin{minipage}[t]{0.42\linewidth}
  \begin{subfigure}[t]{1\linewidth}
  \centering
  \vspace{0mm}      %
  \begin{minipage}{0.84\linewidth}
    \begin{lstlisting}
select R.A
from R,S,T
where R.B-S.B>T.B
\end{lstlisting}
\end{minipage}
    \vspace{-6mm}
    \caption{}
    \label{SQL_Intentional_relations_minus1}
  \end{subfigure}
  \begin{subfigure}[t]{1\linewidth}
  \centering
  \vspace{0mm}
  \begin{minipage}{0.84\linewidth}
    \begin{lstlisting}
select R.A
from R,S,T,">","-"
where R.B="-".left
and S.B="-".right
and ">".left="-".out
and ">".right=T.B
\end{lstlisting}
\end{minipage}
    \vspace{-6mm}
    \caption{named perspective}
    \label{SQL_Intentional_relations_minus2}
  \end{subfigure}
  \begin{subfigure}[t]{1\linewidth}
  \centering
  \vspace{0mm}
  \begin{minipage}{0.84\linewidth}
    \begin{lstlisting}
select R.A
from R,S,T
where "-"(R.B,S.B,x)
and ">"(x,T.B)
\end{lstlisting}
\end{minipage}
    \vspace{-6mm}
    \caption{positional perspective}
    \label{SQL_Intentional_relations_minus3}
  \end{subfigure}
\end{minipage}
\hspace{2mm}
\begin{minipage}[t]{0.53\linewidth}
  \vspace{0mm}    %
  \begin{subfigure}[t]{\linewidth}
    \vspace{3mm}
    \centering        
    \includegraphics[scale=0.5]{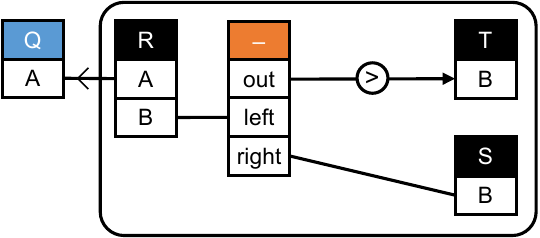}
    \caption{diagram for \cref{trc:minus2}}
    \label{Fig_Intentional_relations_minus1}
\end{subfigure}
\begin{subfigure}[t]{\linewidth}
    \vspace{4mm}  
    \centering        
    \includegraphics[scale=0.5]{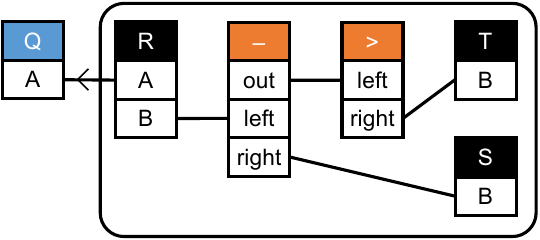}
    \caption{diagram for \cref{SQL_Intentional_relations_minus2}}
    \label{Fig_Intentional_relations_minus2}
\end{subfigure}
\end{minipage}

\caption{In relational languages, 
every computable relation can be relationalized as an \HL{external relation} with externally defined semantics.
For example, (b)-(e) use an external relation for minus called ``$-$''.
(b)/(c): Compare the syntax if external relations are interpreted either under a named or unnamed (positional) perspective
in \SQL.
}
\end{figure}

\begin{example}[arithmetic and comparison operators]
  \label{ex:arithmetic}
A relational interpretation of the arithmetic operator ``Minus'' (for ``$-$'' as in $5-3=2$)
is given by:
\begin{align*}
  \{\textit{Minus}(\textit{left},\textit{right},\textit{out}) \mid \textit{Minus.out} = \textit{Minus.left} - \textit{Minus.right}\}
\end{align*}
Thus, in the following query with an arithmetic minus operator
\begin{align}
  \{Q&(A) \mid 
  \exists r \in R, s \in S, t \in T [Q.A=r.A \wedge r.B - s.B > t.B] \} \label{trc:minus1}
\end{align}
we can \HL{relationalize} the minus operator
(i.e., reify it as a relation) and rewrite the query.
This yields a join query:
\begin{align}
  \{Q&(A) \mid 
  \exists r \in R, s \in S, t \in T, f \in \textit{Minus} [Q.A=r.A \wedge  \label{trc:minus2}\\
  &f.\textit{left}=r.B \wedge f.\textit{right}=s.B \wedge f.\textit{out}>t.B] \}     \notag
\end{align}
We can also \HL{relationalize} the comparison operator $>$ as a separate relation named ``Bigger'':
\begin{align*}
  &\{\textit{Bigger}(\textit{left},\textit{right}) \mid 
  \textit{Bigger.left} > \textit{Bigger.right}\}
\end{align*}
and rewrite the query as an equijoin between relations~\cite{DBLP:journals/pvldb/TziavelisGR21}:
\begin{align}
\{Q&(A) \mid 
  \exists r \in R, s \in S, t \in T, f \in \textit{Minus}, g \in \textit{Bigger}     \label{trc:minus3}\\
  &[Q.A=r.A \wedge f.\textit{left}=r.B \wedge f.\textit{right}=s.B \, \wedge      \notag \\
  &f.\textit{out}=g.\textit{left} \wedge g.\textit{right} = t.B] \}   \notag
\end{align}
Queries \cref{trc:minus1} and \cref{trc:minus3} correspond to the \SQL queries from 
\cref{SQL_Intentional_relations_minus1} 
and
\cref{SQL_Intentional_relations_minus2}, respectively. 
\Cref{Fig_Intentional_relations_minus1,Fig_Intentional_relations_minus2}
show the queries from 
\cref{trc:minus2} and \cref{trc:minus3}, respectively (with minus shown as ``$-$'', etc.).
\end{example}

\begin{figure*}[t]
\centering
\begin{minipage}[t]{0.35\linewidth}
  \vspace{1.8mm}      %
  \begin{subfigure}[t]{1\linewidth}
  \centering        
  \includegraphics[scale=0.4]{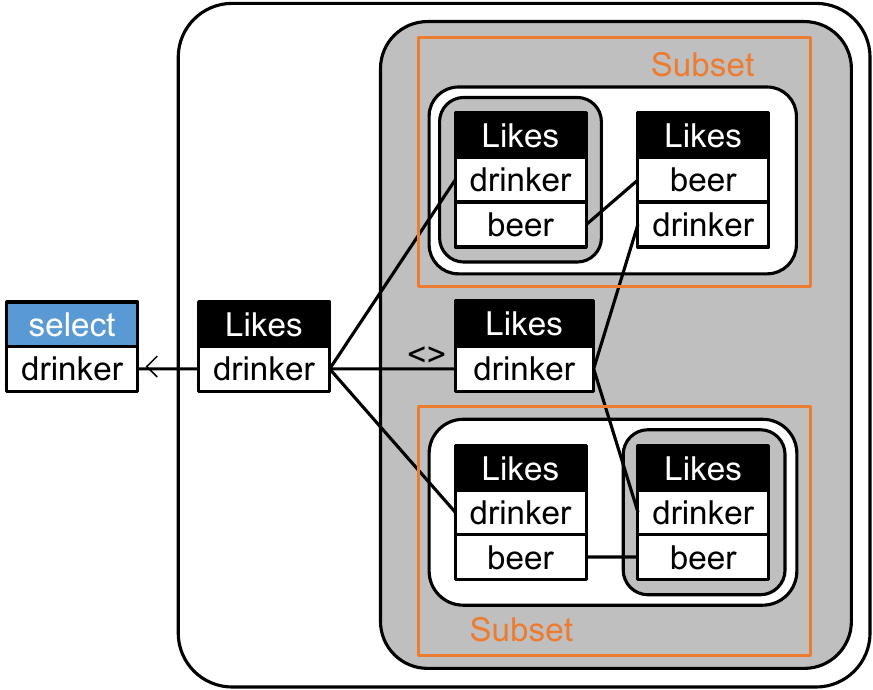}
  \caption{}
  \label{Fig_unique_set_query1}
  \end{subfigure}
\end{minipage}
\hspace{2mm}
\begin{minipage}[t]{0.25\linewidth}
  \vspace{0mm}      %
  \begin{subfigure}[t]{1\linewidth}
  \centering        
  \includegraphics[scale=0.4]{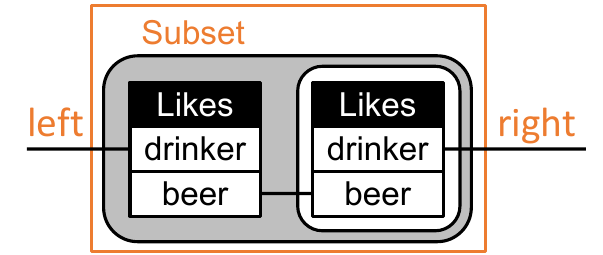}
  \caption{}
  \label{Fig_unique_set_query2}
  \end{subfigure}
  \begin{subfigure}[t]{1\linewidth}
  \centering        
  \includegraphics[scale=0.4]{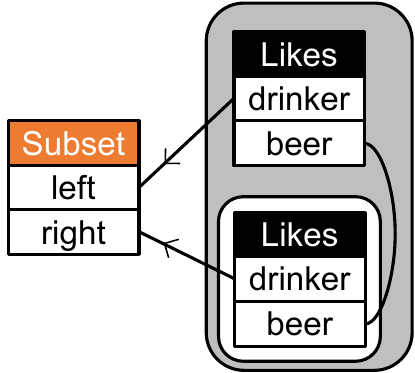}
  \caption{}
  \label{Fig_unique_set_query3}
  \end{subfigure}
\end{minipage}
\hspace{2mm}
\begin{minipage}[t]{0.28\linewidth}
  \vspace{6mm}      %
  \begin{subfigure}[t]{1\linewidth}
  \centering        
  \includegraphics[scale=0.4]{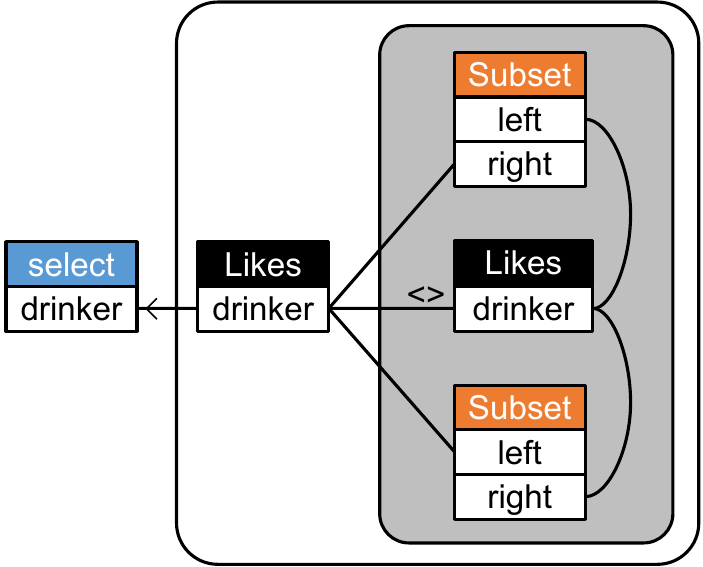}
  \vspace{4mm}    
  \caption{}
  \label{Fig_unique_set_query4}
  \end{subfigure}
\end{minipage}

\caption{By using abstract relations, the unique-set query \cref{SQL_unique-set_query} can be modularized and now more easily interpreted as finding drinkers s.t.\ there is no other drinker 
who likes both a subset and a superset of the beers.
Notice that the newly defined relation ``Subset'' does not have a well-defined extension outside the context in which it is used, and that is OK.
}
\label{Fig:modualarizing the set query}
\end{figure*}

\introparagraph{Discussion}
\circled{1} 
In \cref{ex:arithmetic}, the relational definition of the Minus relation uses the arithmetic operator ``$-$''.
Its meaning is therefore not determined by pure relational operators and must be provided by primitives outside the relational core.
Such primitives are often called ``built-ins''; we use the term \emph{external relation} to emphasize that their semantics is derived from concepts outside core relational constructs.
For example,
$\llbracket \textit{Minus} \rrbracket = \{(x,y,z) \mid z = x - y\}$.
If \sql{Add} is already defined as a primitive operator, then subtraction can also be characterized via addition:
$\llbracket \textit{Minus} \rrbracket = \{(x,y,z) \mid \textit{Add}(y,z,x)\}$.
More generally, we can \HA{relationalize}[relationalize] (reify) such operations
(i.e., treat them as relations) 
to make their use explicit in queries.
We also note that \cref{SQL_Intentional_relations_minus3} illustrates a mixing of the named perspective (SQL)
with an unnamed perspective where the operands of predicates are accessed positionally;
such mixing can break compositionality (here, the join attribute $x$ is not defined).
The formalities of such external specifications are not our focus; we instead focus on modular building blocks of relational languages, of which external relations are one.

\circled{2}
The comprehension-style definition of \textit{Minus} in \cref{ex:arithmetic} also raises the usual safety issue:
none of its ``attributes'' are range-restricted, so the relation is unsafe and ill-defined.
We can restore safety by guarding the operands and result with a domain relation $D(v)$:
\begin{align*}
  &\{\textit{Minus}(\textit{left},\textit{right},\textit{out}) \mid 
  \exists d_1 \in D, d_2 \in D, d_3 \in D [\textit{Minus.left} = d_1.v \wedge           \\
  &\textit{Minus.right} = d_2.v \wedge   
  \textit{Minus.out} =  d_3.v \wedge
  d_3.v = d_1.v - d_2.v  ]\}
\end{align*}

\noindent
For our purpose, we abstract from such concrete definitions: we assume that external relations can be defined meaningfully and are then accessible to the language.
Their concrete realizations are language-specific and not our focus.
Under this abstraction, any computation (including arithmetic operators) can be seen as a relation, and an abstract relational query language treats computation uniformly as relations.

\circled{3} 
Closely related recent work~\cite{DBLP:conf/icdt/GuagliardoLMMMP25}
formalizes external predicates (which evaluate to either true or false when all operands are fixed)
as possibly infinite relations whose extensions are not stored in the database, 
but are accessed through specific \emph{access patterns}.\footnote{
Notice a slightly different motivation for the word ``external predicate'':
For \cite{DBLP:conf/icdt/GuagliardoLMMMP25}, external predicates are ``computed on demand rather than stored'', 
they are \emph{external} to the database,
and only usable through a controlled interface (access patterns).
We use the word external to focus on the fact that their formal definition needs to bring in concepts that are external to the relational model and cannot be described by standard relational operators. 
Both interpretations agree that operations can be \HA{relationalized}[relationalize] 
(reified, i.e.\ turned into relations) for the purpose of analyzing and describing queries.}
Intuitively, a predicate's truth value is a function of its inputs
(e.g., $\textit{Add}(2,3,5)$ is true).
\HA{Access patterns}[access patterns] turn such Boolean predicates into a family of (multi-valued) functions 
that a query engine can call when only a subset of the inputs are fixed (e.g., $\textit{Add}(2,x,5)$ represents $5-2$ and returns $x=3$), 
while still fulfilling safety requirements.
This lets operands of external predicates be joined 
(e.g., the join ``-''.out $=$ ``>''.left between two external predicates in \cref{Fig_Intentional_relations_minus2} prevents them from being evaluated as independent Boolean predicates) and also enables 
such predicates to produce outputs when connected via \HL{assignment predicates}[assignment predicate] 
(see e.g., \cref{sec:matrix muliplication} and \cref{Fig_rel_1}). 
In other words, richer safety conditions allow these predicates to be treated like ordinary database relations
during query evaluation.

\circled{4} 
Other recent work \cite{disjunctions}
\HL{relationalizes}[relationalize] 
join and selection predicates into ``anchor relations''
in order to support arbitrarily nested disjunctions in a diagrammatic presentation (higraph modality).

\subsubsection{\HA{Abstract Relations}[abstract relation]}
\label{sec:abstract relations}
Abstract relations are relation symbols defined \emph{within} a relational language to name and abstract a subquery.
In contrast to \emph{external relations}, abstract relations need not denote a
standalone, well-defined extension on their own.
In particular, an abstract relation may be domain-dependent 
and thus may not have a well-defined extension on its own.
Nevertheless, when an abstract relation occurs inside a safe surrounding query, it can be interpreted as denoting
\emph{some} reasonable finite relation that makes the overall query well-defined.
This is exactly the point of abstraction: when analyzing the intent of the larger query, we do not need to reason about
the internal details of the module or its standalone extension.

\begin{example}[unique-set query]
  \label{ex:unique-set query}
We are given a single relation
\sql{Likes(drinker, beer)}, 
which we abbreviate by $L(d,b)$, and wish to find drinkers who like a unique set of beers, 
i.e., no other drinker likes the exact same set of beers (see \cite[Fig.~1]{Leventidis2020QueryVis},
\cite[Fig.~9]{Gatterbauer2022PrinciplesQueryVisualization} for extensive discussion of this query).
\cref{TRC:subset query}
In the relationally complete fragment (the first-order fragment), the query is written as \cref{SQL_unique-set_query} in \SQL and as follows in \TRC{} and thus also in \ARC:
\begin{align}
& \{Q(d) \mid \exists  l_1 \in L [Q.d=l_1.d \wedge              \label{TRC:subset query}\\
&\quad    \neg (\exists l_2 \in L [l_2.d <> l_1.d \wedge                                  \notag\\ 
&\quad\quad    \neg(\exists l_3 \in L [l_3.d=l_2.d \wedge                                 \notag\\
&\quad\quad\quad    \neg(\exists l_4 \in L [l_4.b=l_3.b \wedge l_4.d=l_1.d]) ]) \wedge    \notag\\
&\quad\quad         \neg(\exists l_5 \in L [l_5.d=l_1.d \wedge                            \notag\\
&\quad\quad\quad    \neg(\exists l_6 \in L [l_6.d=l_2.d \wedge l_6.b=l_5.b]) ]) ]) ] \}   \notag
\end{align}
To modularize the query, we define an abstract relation Subset (denoted $S$ in \ARC):
\begin{align}
& \{S(\sql{left},\sql{right}) \mid                               \label{TRC:subset of subset}\\
&\quad    \neg(\exists l_3 \in L [l_3.d=S.\sql{left} \wedge                             \notag\\
&\quad\quad    \neg(\exists l_4 \in L [l_4.b=l_3.b \wedge l_4.d=S.\sql{right}]) ]) \}   \notag
\end{align}
Taken in isolation, that definition is not safe and therefore does not define a view with a well-defined extension.
But in the context of the enclosing query,
the module represents exactly the intended subset relation between drinkers and allows us to modularize and
compartmentalize the query.
Abstracting it as such we can use it and rewrite the original query more concisely as:
\begin{align}
& \{Q(d) \mid \exists  l_1 \in L [Q.d=l_1.d \wedge              \label{TRC:subset query collapsed}\\
& \quad    \neg (\exists l_2 \in L, s_1 \in S, s_2 \in S [l_2.d <> l_1.d \wedge   \notag \\ 
&\quad    s_1.\sql{left}=l_1.d \wedge s_1.\sql{right}=l_2.d \wedge                \notag \\
&\quad    s_2.\sql{left}=l_2.d \wedge s_2.\sql{right}=l_1.d]) ] \}                  \notag
\end{align}
\end{example}

In the diagrammatic modality, abstract relations correspond to sub-diagrams that can be \emph{collapsed} and \emph{expanded}:
a complex substructure (including its internal scopes) can be replaced by a clearly distinguished module node labeled with
the abstract relation name, and later expanded again.
This supports complexity management for large queries via modularization, hierarchy, and ``zooming''~(see also
\cite[Sect.~4.4]{DBLP:journals/tse/Moody09}).

\begin{figure}[t]
\centering
\begin{minipage}{0.62\linewidth}
\begin{lstlisting}
select distinct L1.drinker
from Likes L1
where not exists
  (select	1
  from Likes L2
  where L1.drinker <> L2.drinker
  and not exists 
    (select 1
    from Likes L3
    where L3.drinker = L2.drinker
    and not exists
      (select 1
      from Likes L4
      where L4.drinker = L1.drinker
      and L4.beer = L3.beer))  
  and not exists
    (select 1
    from Likes L5
    where L5. drinker = L1.drinker
    and not exists
      (select 1
      from Likes L6
      where L6.drinker = L2.drinker
      and L6.beer = L5.beer)))
\end{lstlisting}
\end{minipage}
\vspace{-4mm}
\caption{Unique-set query (\cref{ex:unique-set query}).}
\label{SQL_unique-set_query}
\end{figure}

\begin{figure}[t]
\centering
\begin{subfigure}[t]{1\linewidth}
  \centering
  \begin{minipage}{0.62\linewidth}
\begin{lstlisting}
select distinct D1.drinker as left,
                D2.drinker as right
into Subset
from Likes D1, Likes D2 
where not exists
  (select 1
  from Likes L3
  where not exists
    (select 1
    from Likes L4
    where L4.beer = L3.beer
    and D2.drinker = L4.drinker)  
  and D1.drinker = L3.drinker)  
\end{lstlisting}
  \end{minipage}
  \vspace{-4mm}
  \caption{}
  \label{SQL_unique-set_query_Subset}
\end{subfigure}

\begin{subfigure}[t]{1\linewidth}
  \centering        
  \includegraphics[scale=0.4]{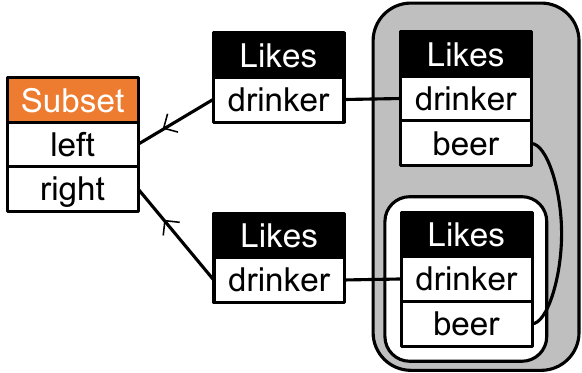}
  \caption{}
  \label{Fig_unique_set_query5}
\end{subfigure}
\caption{Safely defined Subset relation for (\cref{ex:unique-set query}).}
\label{Both_unique_set_subset}
\end{figure}

\begin{figure}[t]
\centering
\begin{minipage}{0.62\linewidth}
\begin{lstlisting}
select distinct L1.drinker
from Likes L1
where not exists
  (select 1
  from Likes L2,Subset S1,Subset S2
  where L1.drinker <> L2.drinker
  and S1.left=L1.drinker 
  and S1.right=L2.drinker
  and S2.left=L2.drinker 
  and S2.right=L1.drinker)
\end{lstlisting}
\end{minipage}
\vspace{-4mm}
\caption{Query from (\cref{ex:unique-set query}) rewritten to use the view from \cref{Both_unique_set_subset}.}
\label{SQL_unique-set_query_New query}
\end{figure}

\section{Two examples}
\label{sec:use cases}

\begin{figure}[t]
    \centering        
    \includegraphics[scale=0.5]{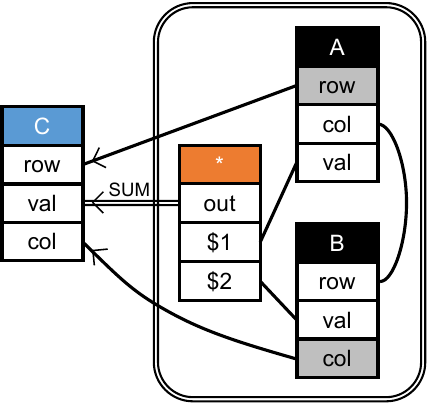}
    \caption{Matrix multiplication \cref{query:rel:matrix multiply}/\cref{eq: rel GTRC 2} in the higraph modality.}
    \label{Fig_rel_1}
\end{figure}

\begin{figure*}[t]
\centering
\begin{subfigure}[b]{.3\linewidth}
  \centering
\begin{minipage}{0.57\linewidth}
\begin{lstlisting}
select R.id
from R 
where R.q = 
  (select count(S.d)
  from S
  where S.id = R.id)
\end{lstlisting}
\end{minipage}
\vspace{-7mm}
\caption{Count bug: version 1}
\label{Fig_countbug_SQL_v1}
\end{subfigure}
\begin{subfigure}[b]{.33\linewidth}
\centering
\begin{minipage}{0.82\linewidth}
\begin{lstlisting}
select R.id
from R, 
  (select S.id, count(S.d) as ct 
  from S
  group by S.id) as X
where R.q = X.ct and R.id = X.id
\end{lstlisting}
\end{minipage}
\vspace{-7mm}
\caption{Count bug: version 2}
\label{Fig_countbug_SQL_v2}
\end{subfigure}
\begin{subfigure}[b]{.34\linewidth}
\centering
\begin{minipage}{0.81\linewidth}
\begin{lstlisting}
select R.id 
from R, 
  (select R2.id, count(S.d) as ct 
  from R R2 left join S
  on R2.id = S.id
  group by R2.id) as X 
where R.q = X.ct and R.id = X.id
\end{lstlisting}
\end{minipage}
\vspace{-7mm}
\caption{Count bug: version 3}
\label{Fig_countbug_SQL_v3}
\end{subfigure}
\begin{subfigure}[b]{.3\linewidth}
  \centering        
  \includegraphics[scale=0.5]{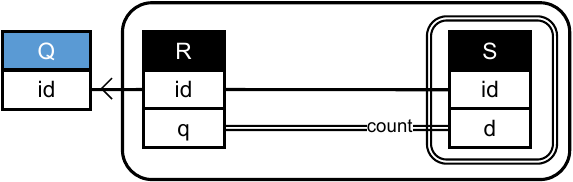}
  \vspace{-1mm}
  \caption{Count bug: version 1}
  \label{Fig_countbug_v1}
\end{subfigure}
\begin{subfigure}[b]{.33\linewidth}
  \centering        
  \includegraphics[scale=0.5]{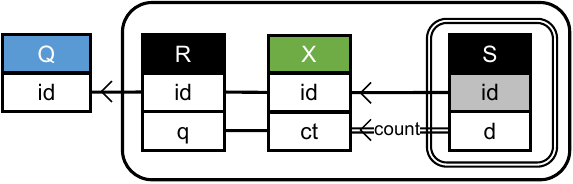}
  \vspace{-1mm}
  \caption{Count bug: version 2}
  \label{Fig_countbug_v2}
\end{subfigure}
\begin{subfigure}[b]{.34\linewidth}
    \centering        
    \vspace{1mm}
    \includegraphics[scale=0.5]{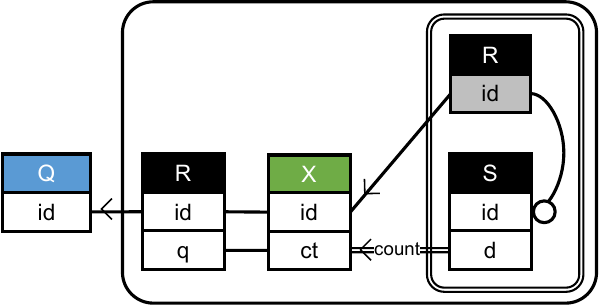}
    \vspace{-1mm}
    \caption{Count bug: version 3}
    \label{Fig_countbug_v3}
\end{subfigure}
\begin{subfigure}[b]{.3\linewidth}
    \centering        
    \includegraphics[scale=0.5]{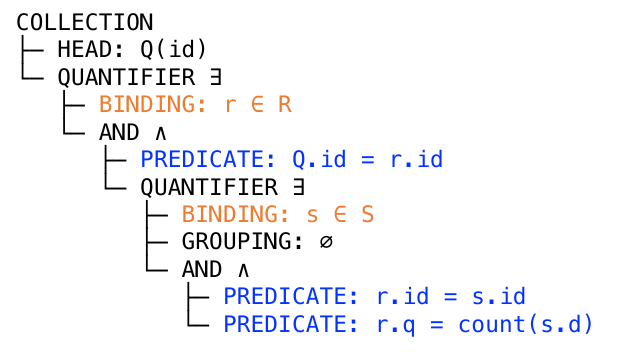}
    \vspace{-3mm}
    \caption{ALT version 1}
    \label{Fig_countbug_ALT_v1}
\end{subfigure}
\begin{subfigure}[b]{.33\linewidth}
    \centering        
    \includegraphics[scale=0.5]{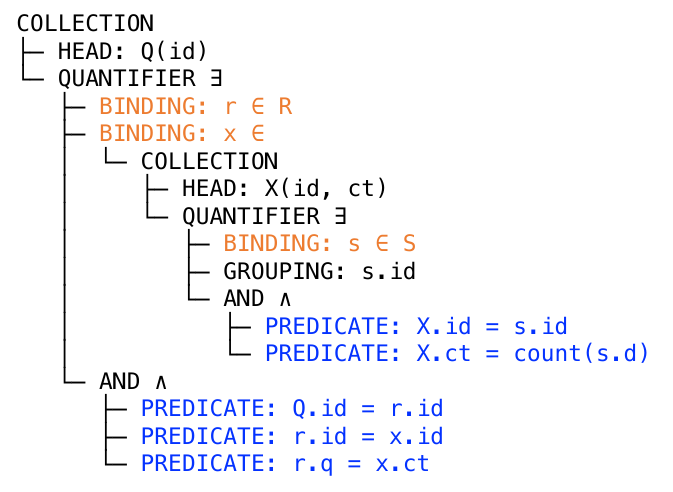}
    \vspace{-3mm}
    \caption{ALT version 2}
    \label{Fig_countbug_ALT_v2}
\end{subfigure}
\begin{subfigure}[b]{.34\linewidth}
    \centering        
    \vspace{1mm}
    \includegraphics[scale=0.5]{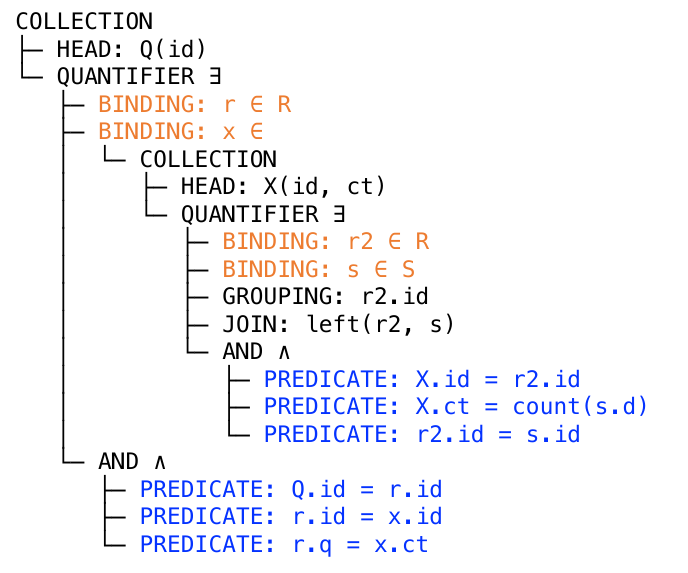}
    \vspace{-3mm}
    \caption{ALT version 3}
    \label{Fig_countbug_ALT_v3}
\end{subfigure}
\caption{\Cref{sec:count bug}: Illustrations of the count bug: 
Left/middle/right 
columns correspond to queries
\cref{gtrc: count bug 1}/\cref{gtrc: count bug 2}/\cref{gtrc: count bug 3}, respectively.
}
\label{Fig:countbug:large}
\end{figure*}

\subsection{Matrix multiplication}
\label{sec:matrix muliplication}

We now illustrate the matrix-multiplication example from the Rel paper~\cite{DBLP:conf/sigmod/ArefGKLMMMMNPRS25}
Rel expresses matrix multiplication between matrices $A$ and $B$ in sparse relational form 
and domain-based positional notation as follows:
\begin{align}
\texttt{def } 
& \texttt{MatrixMult[i,j] :}     \label{query:rel:matrix multiply}  \\[-1mm]
& \texttt{sum[[k] : A[i,k]*B[k,j]]}      \notag 
\end{align}

\noindent
If we allow arithmetic operations in \ARC{} 
and assume all matrices use the same schema 
$(\textit{row},\textit{col},\textit{val})$,
the same computation can be written in the named perspective as:
\begin{align}
  \{C&(\textit{row},\textit{col},\textit{val}) \mid 
  \exists a \in A, b \in B, \gamma_{a.\textit{row}, b.\textit{col}} \notag \\
  &[C.\textit{row}=a.\textit{row}
  \wedge C.\textit{col}=b.\textit{col} 
  \wedge a.\textit{col}=b.\textit{row} \, \wedge            \notag \\
  &C.val = \sql{sum}(a.\textit{val}*b.\textit{val}) ] \}     \notag
\end{align}

\noindent
In the higraph modality~\cref{Fig_rel_1}, multiplication is modeled via an external relation "*"(\$1, \$2, out):
\begin{align}
  \{C&(\textit{row},\textit{col},\textit{val}) \mid 
  \exists a \in A, b \in B, f \in \textrm{"}\!*\!\textrm{"}, \gamma_{a.\textit{row}, b.\textit{col}}    \label{eq: rel GTRC 2}\\
  &[C.\textit{row}=a.\textit{row}
  \wedge C.\textit{col}=b.\textit{col} 
  \wedge a.\textit{col}=b.\textit{row} \, \wedge            \notag \\
  &
  C.val = \sql{sum}(f.\textit{out})
  \wedge f.\$1 = a.\textit{val} 
  \wedge f.\$2 = b.\textit{val} 
  ] \}     \notag
\end{align}

\subsection{An illustration of the count bug}
\label{sec:count bug}

The count bug~\cite{DBLP:conf/sigmod/GanskiW87} is a famous example of an attempted reformulation of a nested correlated query 
like the one in \cref{Fig_countbug_SQL_v1} and replacing it with
\cref{Fig_countbug_SQL_v2}
(the incorrect translation was given in \cite{DBLP:journals/tods/Kim82},
and corrected in \cite{DBLP:conf/sigmod/GanskiW87})
However, 
on an input database with $R(9,0)$ and empty table $S$, the first query would return $9$, whereas the second would return an empty result. 
The correct decorrelation happens with left join and the query shown in 
\cref{Fig_countbug_SQL_v3}.\footnote{The example assumes \sql{R.id} is a key. Otherwise, the correct translation requires an additional deduplication.}
The remaining equations and figures in this section show those queries in a pattern-equivalent \ARC{} representations and various modalities.
\begin{align}
  \{Q(\textit{id}) \mid \,      
  &\exists r \in R 
  [Q.\textit{id}=r.\textit{id} 
  \wedge 
  \exists s \in S,
  \gamma_{\emptyset}                    \label{gtrc: count bug 1}
  \\
  &[r.\textit{id}=s.\textit{id}
  \wedge 
  r.q=\countagg(s.d)] ] \} \notag \\
  \{Q(\textit{id}) \mid \, &
  \exists r \in R, x \in \{X(\textit{id}, \textit{ct}) \mid \,  \label{gtrc: count bug 2}
  \\
  &\exists s \in S,
  \gamma_{s.id}
  [X.\textit{id}=s.\textit{id}
  \wedge X.\textit{ct}=\sql{count}(s.d) ] \}    \notag\\
  &[Q.\textit{id}=r.\textit{id}     
  \wedge r.\textit{id}=x.\textit{id}
  \wedge r.q=x.\textit{ct} ] \}     \notag \\
  \{Q(\textit{id}) \mid \, &
  \exists r \in R, x \in \{X(\textit{id}, \textit{ct}) \mid 
  \exists s \in S, r_2 \in R,
  \gamma_{r_2.id}, \textsf{left}(r_2,s)   \label{gtrc: count bug 3} 
  \\
  &[X.\textit{id}=r_2.\textit{id}
  \wedge X.\textit{ct}=\sql{count}(s.d) 
  \wedge r_2.\textit{id}=s.\textit{id} ] \}   \notag \\
  &[Q.\textit{id} = r.\textit{id}
  \wedge r.\textit{id}=x.\textit{id}
  \wedge r.q=x.\textit{ct} ] \}     \notag   
\end{align}

\section{Questions \& Answers}

\emph{\textbf{The flat relational model is obsolete. We need to think bigger}
and move to an Entity-Relational Data model ERDs~\cite{Desphande:CIDR:2025}, 
relational maps~\cite{DBLP:journals/corr/abs-2504-12953}, 
databases as output~\cite{DBLP:journals/pacmmod/NixD25},
or at least a nested relational model~\cite{DBLP:conf/icde/0001CGOPSVW24}}.
While we see merit in these directions, our goal is more modest.
Rather than replace or extend the relational model, 
our goal is to complement current practices.
SQL remains widely used, and when inputs and outputs are in 1NF, 
nesting of intermediate results adds no expressive power~\cite{DBLP:journals/tcs/Libkin03}. 
That said, nested relations are now part of standards~\cite{DBLP:conf/icde/0001CGOPSVW24}
and the development of an abstract nested relational QL remains open.

\emph{\textbf{SQL is bad for users. We need to create a new language},
like SaneQL~\cite{DBLP:conf/cidr/0001L24}, Rel~\cite{DBLP:conf/sigmod/ArefGKLMMMMNPRS25},
or a pipe/dataflow syntax~\cite{DBLP:journals/pvldb/ShuteBBBDKLMMSWWY24}}.
That may or may not be true, but our goal is orthogonal: 
rather than propose yet another QL or extension to SQL, 
we suggest that the database community instead creates a relational \emph{reference language} (a relational \emph{meta language})
that abstracts the relational patterns across query languages (the relational core) 
away from a surface syntax.

Usability is not an immutable property of a language, it also depends on the \emph{modality} in which it is presented.
Therefore, \ARC{} provides multiple modalities, including a machine-oriented data structure called \emph{Abstract Language Tree} (ALT).
Because ALT exposes bindings, scopes, and grouping structure directly, it supports systematic traversal, rewriting, and validation, and can serve as an intermediate representation for NL2SQL systems that translate natural language into query intent and then render to \SQL.
Ultimately, questions about relative usability need to be solved with reproducible, task-oriented user studies~\cite{10.1145/3639316,10.1145/3687998.3717044}.

\emph{\textbf{Is this then another Intermediate Representation?}}
Our goal is fundamentally different from intermediate representations (IRs) like 
Semiring Dictionaries~\cite{DBLP:journals/pacmpl/ShaikhhaHSO22},
Substrait~\cite{substrait},
and relational maps~\cite{DBLP:journals/corr/abs-2504-12953},
which are typically closely tied to execution models, and designed to support optimization.
In contrast, we aim to move in the opposite direction, toward a more abstract representation 
that decouples from both data layout and syntax, focusing instead on the semantic structure of relational queries.

\emph{\textbf{Is this about logical expressiveness with aggregates?}}
While the addition of aggregate functions to logic has traditionally been studied in terms of logical expressiveness~\cite{DBLP:journals/tcs/Libkin03}, 
our goal is different. 
Our focus is to capture relational patterns across various languages.
For that purpose we designed a reference language that
supports aggregate queries with the same relational patterns as SQL, including multiple aggregates evaluated within a single grouping scope.

\emph{\textbf{Are you reinventing Rel?}}
No. We fully embrace the Rel philosophy~\cite{DBLP:conf/sigmod/ArefGKLMMMMNPRS25,rel_cheatsheet}: 
everything is a relation, and some relations are defined (or derived) rather than stored.
Rel aims to unify data modeling, querying, and application logic within a single relational language, 
removing the boundary between a relational query sublanguage (e.g., SQL) 
and the host application language (e.g., Java) so that a powerful execution engine can optimize globally across the entire program.
Our aim is not to design such a language or an execution engine.
Instead, our goal is to support the ongoing discussion about both user-facing and machine-facing language 
design trade-offs, the different patterns that appear between relational languages, and the recurring relational core expressible in all of them.

To this end, we provide
an expressive \emph{abstract relational query language} 
that comes with a pattern-preserving diagrammatic ``modality''.
This language is intended as a \emph{reference language} for
analyzing current and future relational languages,
whether they adopt set or bag semantics (which we regard as a ``convention'' orthogonal to a language itself).

\emph{\textbf{Why are we still talking about new languages if everything will be NL2SQL anyway?}}
\ARC/ALT can be used as an intermediate target: 
models generate a structurally constrained representation, which can be validated (well-scoped variables, grouping legality, correlation shape) and then rendered to \SQL. 
This enables intent-based evaluation and comparison of generated queries at the semantic-structure level rather than at the surface-syntax level.

\emph{\textbf{Do you really expect people to write queries in a complicated looking formalism like \cref{gtrc: count bug 3}?}}
No. Our point is not to replace \SQL, \Datalog, or any other surface syntax. 
The point is to make query meaning sayable.
We propose \emph{a universal relational reference language}, 
paired with a \emph{shared vocabulary}, 
that names the primitive operations by which queries combine relations to answer relational questions.
Once those primitives are named, we can talk about the same underlying patterns across very different languages (declarative, procedural, or functional) without mistaking syntax for substance.
Thus, a reference language must be explicit, and explicitness is often verbose: 
what production languages compress into syntactic sugar, convention, or ``obvious'' readings must be surfaced if we want a reliable reference point for comparison. 
That surfacing is not busywork; it allows us to name distinctions that our current vocabulary blurs. 
It lets us point at a query in Souffl\'e and say ``\HL{FOI} aggregation.''
It lets us look at \cref{Fig_rel_1} and see the relational pattern for matrix multiplication.
And it lets us diagnose bugs by naming the difference between an aggregate used as a value (assignment predicate) 
and an aggregate used as a test (comparison predicate), as in the count bug 
(\cref{Fig:countbug:large}).

And anyway, fewer people will write queries directly in the future;
more people will read them and try to make sense of them.
In that setting, usability is not just the language, it is about the modality in which it is presented. 
The same semantics can be rendered as text, diagrams, or ALTs; different modalities serve different readers. 
For humans, initial evidence suggests that a diagrammatic modality like \cref{Fig_countbug_v3} 
can be read faster and more accurately than SQL~\cite{10.1145/3639316, 10.1145/3687998.3717044}.
For machines (including LLMs), we believe that explicit, modular structure with small, reusable vocabulary 
can improve precision and recall.

\section{Next steps}

We view Abstract Relational Calculus (\ARC) as a candidate \emph{semantic backbone}:
a relational metalanguage for connecting the surface syntax of queries to their core relational intent, 
across query languages, modalities, and conventions.

On the systems side, we are building a \SQL $\leftrightarrow$ \ARC{} translator
that can render \ARC{} in all 3 modalities,
extending our prototype implementation~\cite{DBLP:conf/sigmod/SabaleG25} 
to cover additional aggregation-nesting patterns and disjunctions.
A spring 2026 seminar on relational language design~\cite{7575:2026} focuses on the pattern expressiveness of relational languages and, in the process, produce further embeddings of other languages in \ARC.

On the theory side, we plan to prove coverage results: for a well-defined fragment of \SQL 
(including arbitrarily correlated queries and aggregation patterns), 
every query has a pattern-preserving \ARC{} representation,
and thus \SQL $\leftrightarrow$ \ARC{} round-tripping is semantics-preserving
(with appropriate conventions).

Finally, an open question is whether \ARC/ALT can indeed serve as an effective intermediate target for NL2SQL, together with datasets and evaluation metrics that score intent via semantic structure proxies (scopes, joins, relational patterns) rather than \SQL syntax similarity.
Also open are extensions to sorted lists (ORDER BY),
as well as extensions to the nested relational model~\cite{DBLP:conf/icde/0001CGOPSVW24}.

\section{Acknowledgements}

We thank 
Molham Aref~\cite{DBLP:conf/sigmod/ArefGKLMMMMNPRS25},
Torsten Grust~\cite{DBLP:journals/jiis/GrustS99}, 
Leonid Libkin~\cite{DBLP:journals/jacm/HellaLNW01,DBLP:conf/sigmod/ArefGKLMMMMNPRS25}, 
Wim Martens~\cite{DBLP:conf/sigmod/ArefGKLMMMMNPRS25},
Amir Shaikhha~\cite{DBLP:journals/pacmpl/ShaikhhaHSO22}, and
Dan Suciu~\cite{DBLP:journals/sigmod/BunemanLSTW94}, 
for taking the time to discuss their respective papers with us,
and Mahmoud Abo Khamis for helping us understand Rel queries.
Wolfgang was supported in part by the National Science Foundation (NSF) under award IIS-1762268,
and Diandre by the NSF Graduate Research Fellowship Program (GRFP).

\bibliographystyle{ACM-Reference-Format}
\bibliography{queryvis-CIDR.bib}

\end{document}